\documentclass[prl, twocolumn, superscriptaddress]{revtex4}
\usepackage{graphicx}
\usepackage{setspace,transparent}
\usepackage{color, dsfont}
\usepackage{placeins}
\usepackage{fontenc,ragged2e,wrapfig}
\usepackage{mathtools,amssymb,amsmath,nicefrac,mathrsfs}
\usepackage[font=footnotesize]{caption}
\usepackage{soul}
\usepackage{hyperref}
\setcitestyle{super}
\usepackage[dvipsnames]{xcolor}
\usepackage[utf8]{inputenc}
\usepackage{tikz, tikz-cd}
\usepackage{listings}
\usepackage[paperwidth=220mm, paperheight=290mm, centering, hmargin=2.0cm, vmargin=2.0cm]{geometry}

\DeclareMathAlphabet{\mathpzc}{OT1}{pzc}{m}{it}
\DeclareCaptionJustification{justified}{\justifying}
\DeclareCaptionJustification{justified}{\justifying}
\DeclareMathAlphabet{\mathpzc}{OT1}{pzc}{m}{it}

\usepackage[LGR,T1]{fontenc}
\newcommand\Koppa{\begingroup\fontencoding{LGR}\selectfont\char21\endgroup}

\begin{document}
\title{Hybridized universality classes of long-range cascades}
\title{Hybrid spinodals for long-range cascades}

\author{\textsc{I.\ Bonamassa}}
\email{ivan.bms.2011@gmail.com, bonamassai@ceu.edu}
\affiliation{Department of Network and Data Science, CEU, Quellenstrasse 51, A-1100 Vienna, Austria}
\author{\textsc{B.\ Gross}}
\affiliation{Network Science Institute, Northeastern University, Boston, MA 02115, USA}
\author{\textsc{J.\ Kert\'esz}}
\affiliation{Department of Network and Data Science, CEU, Quellenstrasse 51, A-1100 Vienna, Austria}
\author{\textsc{S.\ Havlin}}
\affiliation{Department of Physics, Bar-Ilan University, 52900 Ramat-Gan, Israel}

\date{\today}

\begin{abstract} 
Cascades are self-reinforcing processes underlying the systemic risk of many complex systems. Understanding the universal aspects of these phenomena is of fundamental interest, yet typically bound to numerical observations in ad-hoc models and limited insights. Here, we develop a unifying approach and show that cascades induced by a long-range propagation of local perturbations are characterized by two universality classes determined by the parity invariance of the underlying process. We provide hyperscaling arguments predicting hybrid critical exponents given by a combination of both mean-field spinodal exponents and $d$-dimensional corrections and we show how global symmetries influence the geometry and lifetime of avalanches. Simulations encompassing classic and novel cascade models validate our predictions, revealing fundamental principles of cascade phenomena amenable to experimental validation. 
\end{abstract}

\maketitle

\begin{figure}[h]
	\includegraphics[width=0.95\linewidth]{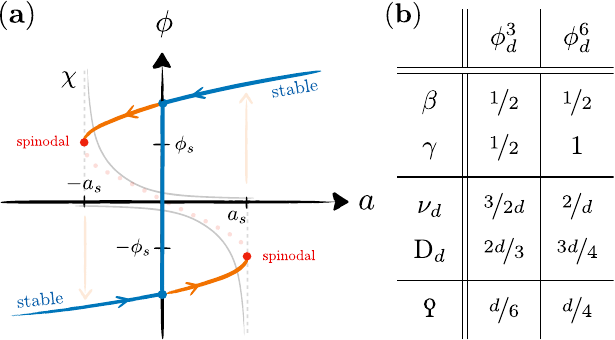}
		\caption{\footnotesize{\textbf{Spinodals \& long-range cascades.} 
		(\textbf{a})~In first-order transitions, the spinodal singularity, $a_s$ (red circle), marks the limit of metastability (orange curves) beyond the coexistence threshold (blue symbol). When the control parameter $a\to \pm a_s^\mp$, the critical fluctuations of the order parameter $\phi$ diverge, producing a one-sided susceptibility $\chi\propto |a-a_s^\pm|^{-\gamma}$ (grey curves). 
		(\textbf{b})~Classes of long-range cascades: strong scaling exponents, like $\beta$, $\gamma$ or $\tau$, match spinodal ones in $\phi^3$ and $\phi^6$ mean-field theories; weak scaling exponents, like $\nu$, $\mathrm{D}$ or $\text{\Koppa}$, depend on the dimension $d$. 
		}}\vspace*{-0.25cm}
		\label{fig:1}
\end{figure}

Dependency couplings, load sharing~\cite{motter2017unfolding} and other positive feedback mechanisms~\cite{deangelis2012positive} often cause a non-linear response of a system's state to local perturbations, amplifying their influence up to length scales determined by the range of the interactions of the system. 
Long-range interactions, in particular, ignite cascades that can propagate {\em at all} scales~\cite{dodds2004universal, haldane2011systemic, barnosky2012approaching, brummitt2015coupled, majdandzic2016multiple, schafer2018dynamically, rocha2018cascading, artime2020abrupt, scheffer2020critical}, yielding first order transitions with scaling $\phi(a)-\phi_s\propto |a-a_s|^{\beta}$ and other critical signatures (Fig.~\ref{fig:1}\textbf{a}). These intriguing transitions, often called mixed-order because of the simultaneous presence of criticality and of an abrupt jump~\cite{boccaletti2016explosive, d2019explosive, kuehn2021universal}, have been reported for cascading processes on both random and spatial networks~\cite{gross2023dynamics}, emphasizing their long-range kinetics. Examples include, but are not limited to, interdependent percolation~\cite{buldyrev-nature2010, bashan2013extreme}, higher-order dynamics~\cite{iacopini2019simplicial, battiston2021physics, zhang2023higher, ferraz2023multistability}, traffic~\cite{zeng2020multiple} or flow redistribution~\cite{motter2002cascade,hoffmann2014suppressing,zhao2016spatio}, jammed packings~\cite{henkes2005jamming, silbert2005vibrations, schwarz2006onset} and fractures by elastic or electric forces~\cite{ferguson1999spinodals, petri1994experimental, moreno2000fracture, rundle2003statistical}.

In this Letter, we show that long-range cascades yielding mixed-order transitions with exponent $\beta=1/2$ can be grouped into two universality classes defined by the parity invariance of the cascading process. By identifying critical cascades as spinodal fluctuations~\cite{heermann1982spinodals, unger1984nucleation, monette1994spinodal, klein2007structure}, we introduce hyperscaling arguments predicting ``hybrid'' critical exponents ---i.e.\ endowed with both mean-field and $d$-dimensional features--- for the correlation length, $\nu_d$, the fractal dimension, $\mathrm{D}_d$, and the anomalous dimension, $\eta_d$, of avalanches in the two classes for {\em any} dimension $d$ of the underlying network (Fig.~\ref{fig:1}\textbf{b}). In parity-breaking processes ---for brevity, percolation or $\phi^3_d$-cascades--- we show that the critical fluctuations accompanying the correlation length divergence, $\xi\propto|a-a_s|^{-\nu_d}$, decay faster, $\nu_d=3/2d$, then statistical ones~\cite{gross2022fractal} ($\nu'_d=2/d$), leaving their signatures hidden under white noise. This latent criticality becomes more delicate in parity-invariant or $\phi^6_d$-cascades, where critical ($\nu_d=2/d$) and statistical fluctuations blend on comparable scales. Based on extensive simulations, we demonstrate that a finite-size analysis of the critical window~\cite{stauffer_book} and of the size and distribution of finite avalanches, 
allows to distinguish between critical contribution to scaling and stochastic ones. Doing so, we validate our predictions in a variety of synthetic and experimentally-driven models and discuss key principles of cascade phenomena stemming from our results. 


\begin{center}
\textsc{Cascades, Spinodals \& Hyperscaling}\vspace*{-0.25cm}
\end{center}
$\,\,\,$ Viewed from the theory of first-order phase transitions~\cite{binder1987theory}, the exponents $\beta,\gamma=1/2$ 
(Fig.~\ref{fig:1}\textbf{b}) measured in various mixed-order transitions~\cite{boccaletti2016explosive, d2019explosive, kuehn2021universal} are hallmarks of the {\em mean-field} spinodal critical point in the $\phi^3$ Landau-Ginzburg theory~\cite{unger1984nucleation, monette1994spinodal}. Indeed, in systems undergoing mixed-order transitions, adaptive feedback set in through long-range (global or randomly distributed) interactions~\cite{gross2023dynamics}, enabling avalanches to spread at all distances in a mean-field fashion, {\em even in low dimensions}. In light of this, we argue here that long-range cascades resulting in mixed-order transitions physically realize spinodal fluctuations~\cite{klein2007structure}, prompting two remarks: 

\begin{enumerate}
\item[$\mathrm{R}_1$.] $\phi^6$-theory~\cite{binder1987theory}, i.e.\ systems with parity invariance, yields spinodals with $\beta=1/2$ {\em but} $\gamma=1$ (Fig.~\ref{fig:1}\textbf{b}), hinting at a second class of long-range cascades; 
\item[$\mathrm{R}_2$.] in mean-field regimes, finite-size and hyperscaling relations are notoriously subtle~\cite{aizenman1997number, coniglio1985shapes, binder1985finite, binder1985critical, luijten1996finite, parisi1996scaling, berche2012hyperscaling, kenna2014fisher, kenna2017universal, berche2022phase}, posing equal issues for the study of $\phi^3_d$ and $\phi^6_d$ cascades. 
\end{enumerate}

The latter is a crucial point for our study. 
First, recall that hyperscaling relations like $2\beta+\gamma=d\nu$ and finite-size scaling (FSS) hold for systems below their upper critical dimension $d_c$, where the ratio $\xi/L$ governs FSS behaviors~\cite{fisher1967theory}. 
For $d\geq d_c$, i.e.\ in mean-field regimes, this picture was long considered to fail due to new diverging length scales coming in play. This motivated \textsc{Binder} {\em et al.}~\cite{binder1985finite, binder1985critical} to introduce a ``thermodynamic'' length, $\ell\propto|a-a_s|^{-\nu_T}$ with $\nu_T\equiv \nu_{\mathrm{mf}}d_c/d$, governing FSS via $\ell/L$. Elevating the latter to $d$, one can read this equivalently by saying that, above $d_c$, FSS holds if the mean-field correlated volume $\xi^{d_c}$ scales with the available volume $L^d$. i.e.\ FSS is realized through the ratio $\xi/L^{\text{\Koppa}}$ where $\text{\Koppa}\equiv d/d_c$. Elaborating on this argument, \textsc{Kenna} {\em et al.}~\cite{berche2012hyperscaling, kenna2014fisher, kenna2017universal, berche2022phase} recently developed a \textsc{RG} framework for continuous transitions above $d_c$ resolving, among other puzzles, hyperscaling. These advances provide a reference to understand the rather unconventional critical properties of long-range cascades. 

Let us return to our hypothesis that long-range cascades realize mean-field (spinodal) critical phenomena for any dimension $d$ of the underlying network. Leveraging on the above, this can be understood in two ways. 

$\boldsymbol{i)}$ {\em Each $d$ is an upper critical dimension for the cascading process}, so that classic hyperscaling holds for every $d$. In this portrait, strong-scaling exponents like $\beta,\gamma,\sigma,\dots$ ---i.e.\ those unaffected~\cite{kirkpatrick2015exponent} by dangerous irrelevant variables (DIVs)--- are those of spinodals in $\phi^3$ or $\phi^6$ Landau theory, while weak-scaling exponents depend on $d$. In particular, hyperscaling yields $\nu_d=(2\beta+\gamma)/d$, the fractal dimension of avalanches satisfies the classic relation $\mathrm{D}_d=d-\beta/\nu_d$ and the anomalous dimension (ruling the decay of the correlation function) follows Fisher's relation $\eta_d=2-\gamma/\nu_d$. Thus, long-range cascades can be characterized by the classic spinodal strong-scaling exponents~\cite{monette1994spinodal} and the weak-scaling exponents 
\begin{equation}\label{eq:1}
\!\!\!(\nu_d,\mathrm{D}_d,\eta_d)=\!
\begin{cases}
\big(\nicefrac{3}{2d}, \nicefrac{2d}{3}, 2-\nicefrac{d}{3}\big)&\text{in}\,\,\,\phi^3_d\,\text{-cascades},\\
\big(\nicefrac{2}{d}\,\,, \nicefrac{3d}{4}, 2-\nicefrac{d}{2}\big)&\text{in}\,\,\, \phi^6_d\,\text{-cascades}. 
\end{cases}
\end{equation}
\noindent 
Notice that FSS holds here via the usual length ratio $\xi/L$, where the correlation length $\xi$ coincides with \textsc{Binder}'s thermodynamic length $\ell$, {\em for any} dimension $d$.

$\boldsymbol{ii})$ Both weak {\em and} strong scaling exponents of cascades are, {\em tout court}, those of classic spinodals but FSS emerges through the ratio $\xi /L^{\text{\Koppa}}$ {\em for any $d$}. In this case, \textsc{Kenna} {\em et al.}\ relations~\cite{kenna2014fisher, kenna2017universal} {\em hold both above and below the spinodal's upper critical dimension} $d_c$, so that $\mathrm{D}_d=\text{\Koppa}(\beta+\gamma)/\nu$, with $\nu=1/4$ ($\nu=1/2$) for $\phi^3_d$ ($\phi^6_d$) cascades, and $\eta_d=2(1-\text{\Koppa})$, whose values coincide with those in Eq.~\eqref{eq:1}. 

It is worth stressing that the two scenarios above are equivalent: while the former includes corrections to DIVs into the weak-scaling critical exponents, the latter introduces the exponent $\text{\Koppa}\equiv d/d_c$ to correct FSS. Notably, both scaling pictures hold {\em for every dimension $d$.} That is, the long-range nature of cascades in $d$-dimensional networks emerges through a bland of mean-field spinodal exponents and $d$-dimensional DIVs corrections, motivating the name ``{\em hybrid}\,'' critical phenomena. In the rest of this Letter we verify these predictions in several synthetic and experimentally-driven cascade models. 

\begin{figure}[t]
	\includegraphics[width=\linewidth]{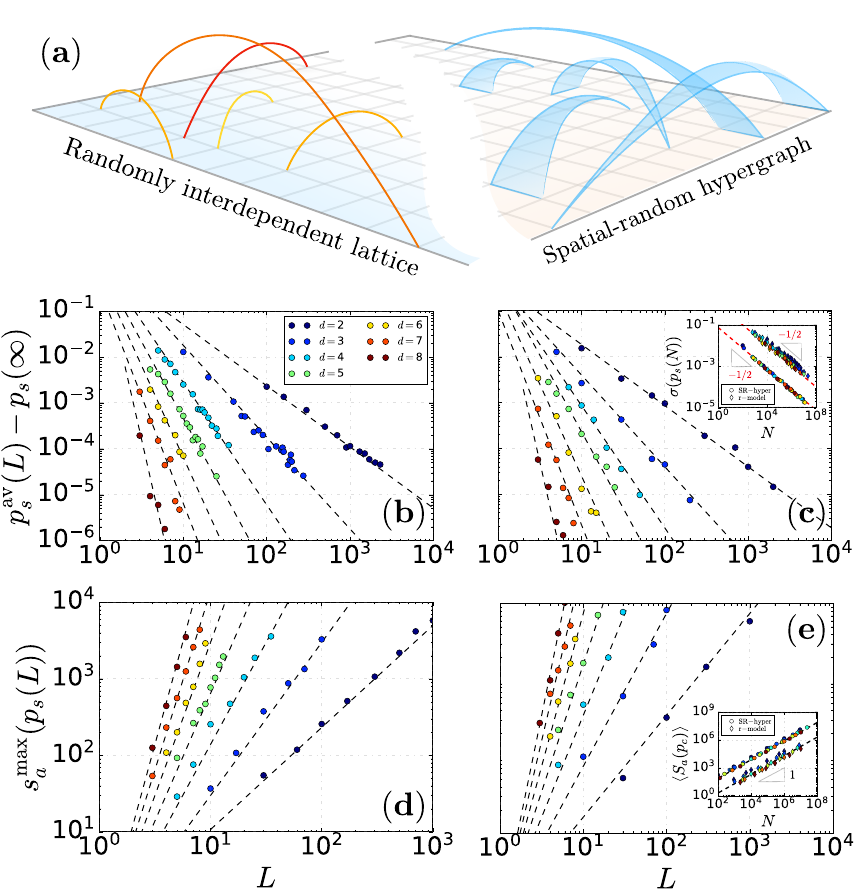}
		\caption{\footnotesize{\textbf{Correlation and geometric exponents.} 
		(\textbf{a}) Illustration of the long-range $\phi^3_d$-cascade models in $d=2$: (left) randomly interdependent lattice and (right) a spatial-random ($d$R) hypergraph (see text for details). 
		(\textbf{b}) Finite-size (FS) scaling of the {\em mean} critical width in interdependent percolation and 
		(\textbf{c}) in $2$-core percolation for increasing dimension, $d$, and linear size, $L$, of the underlying lattice; dashed lines have slopes $-2d/3$ for $d=2,\dots,8$. 
		(Inset, \textbf{c}) Fluctuations of the FS-thresholds, $\sigma(p_s(N))\propto N^{-1/2}$, with $N=L^d$. 
		(\textbf{d}) FS-scaling of the mean size of large finite avalanches sampled at $p_s(L)$ in interdependent percolation and (\textbf{e}) in $2$-core pruning; dashed lines correspond to the fractal dimension $2d/3$ for $d=2,\dots,8$. (Inset, \textbf{e}) Infinite avalanches in both models are compact as they dismantle the whole structure.
}} \vspace*{-0.25cm} 
\label{fig:2}
\end{figure}

\begin{center}
\textsc{Hidden under white noise}\vspace*{-0.25cm}
\end{center}
$\,\,\,$ A striking property stemming from Eq.~\eqref{eq:1} is that the $\nu_d$ exponents (equiv., the $\text{\Koppa}$ exponents) are marginal with respect to Chayes' bound $\nu_d\geq2/d$ (equiv., $\text{\Koppa}\geq d/4$) for finite-size correlations in disordered systems~\cite{chayes1986finite}. That is, unlike other critical phenomena, the critical fluctuations of long-range cascades are hidden under the Gaussian noise generated by disorder. Since the latter dominates FSS at large $L$, some care should be taken when extrapolating $\nu_d$, $\mathrm{D}_d$ or $\eta_d$ via e.g.\ finite-size data collapse~\cite{lee2016hybrid}. 

To circumvent this issue, we study the convergence of the average finite-size (a.k.a.\ pseudocritical) threshold $a_s(L)$ towards its asymptotic value $a_s(\infty)$ as $L$ increases~\cite{stauffer_book}. Notice that this measure is not influenced by statistical fluctuations since $a_s^{\mathrm{av}}(L):=\int a \partial_a\Pi(a,L)\mathrm{d}a$, where $\Pi(a, L)=\Phi[|a-a_s^{\infty}|f(L)]$ is the probability of a mixed-order transition by cascades and $\Phi$ is a scaling function that depends on $L$ via $f(L)=c_1 L^{1/\nu_d}+c_2 L^{1/\nu'_d}$, where $\nu'_d=2/d$ is the contribution of white-noise. Since $\nu'_d\geq \nu_d$, with $\nu_d$ as in Eq.~\eqref{eq:1}, to leading orders 
\begin{equation}\label{eq:2}
\big|a_s^{\mathrm{av}}(L)-a_s^\infty\big|\propto L^{-1/\nu_d}.
\end{equation}
On the other hand, the width $\sigma(a_s(L))$ of the distribution is inversely proportional to $\partial_a\Pi(a, L)$ thus $\sigma\propto1/f(L)\sim L^{-1/\nu'_d}$, explaining the white noise dominance at large $L$. Notably, the latter influences the finite-size averaging and collapse analysis of observables like $\phi(a)$ or $\chi(a)$ but not the finite-size convergence of $a_s^{\mathrm{av}}(L)$ to $a_s^\infty$. Analogous arguments hold in \textsc{Kenna} {\em et al.}~\cite{berche2012hyperscaling, kenna2014fisher, kenna2017universal, berche2022phase} FSS scheme. 

Next, we validate Eq.~\eqref{eq:2} for $\phi^3_d$ long-range cascades in two models of structural dismantling (Fig.~\ref{fig:2}\textbf{a}). We start from interdependent percolation in randomly coupled $d$-dimensional lattices~\cite{bashan2013extreme}, where dependency links randomly couple the percolating state of nodes in one-to-one fashion while retaining the underlying lattice connectivity (Fig.~\ref{fig:2}\textbf{a}, left). As the damage of one node causes the damage of another one at any distance, long-range cascades propagate iteratively after the removal of an initial fraction $1-p$ of randomly selected sites. To ease the computing, we adopt a single layer with randomly distributed dependency links, given its equivalence at criticality with the case of $2$ (or more) randomly coupled lattices. Figure \ref{fig:2}\textbf{b} displays the results testing Eq.~\eqref{eq:2} for increasing dimensions: as visible, the one-parameter family of critical exponents $\nu_d=3/2d$---equivalently, $\nu=1/4$ and $\text{\Koppa}=d/6$---nicely fits the numerical data. 

To consolidate the above results, we introduce a spatial-random ($d$R) hypergraph model constructed by taking each bond of a square lattice in $d$ dimensions as the base of a triangle whose tip is a randomly chosen site (Fig.~\ref{fig:2}\textbf{a}, right). Motivated by the mapping~\cite{bonamassa2021interdependent} of interdependent percolation in $K$ randomly coupled Erd\H{o}s-R\'enyi graphs onto random $(K+1)$-\textsc{xorsat}~\cite{leone2001phase,mezard2003two}, we perform $2$-core percolation on $d$R-hypergraphs. For any $d\geq2$, this process undergoes a mixed-order transition with the critical hallmarks of $\phi^3$ spinodals, i.e.\ $\beta,\gamma=1/2$, and the results shown in Fig.~\ref{fig:2}\textbf{c} show excellent agreement with the correlation exponent $\nu_d=3/2d$ in Eq.~\eqref{eq:2}. The inset of Fig.~\ref{fig:2}\textbf{c} supports the stochastic dominance over critical fluctuations, demonstrated here by analyzing the FSS of the critical width, $\sigma(a_s)\propto L^{-1/\nu'_d}=N^{-1/2}$, corresponding to the data in Fig.~\ref{fig:2}\textbf{b},\,\textbf{c}. 

\begin{figure}[b]\vspace*{-0.5cm}
	\includegraphics[width=\linewidth]{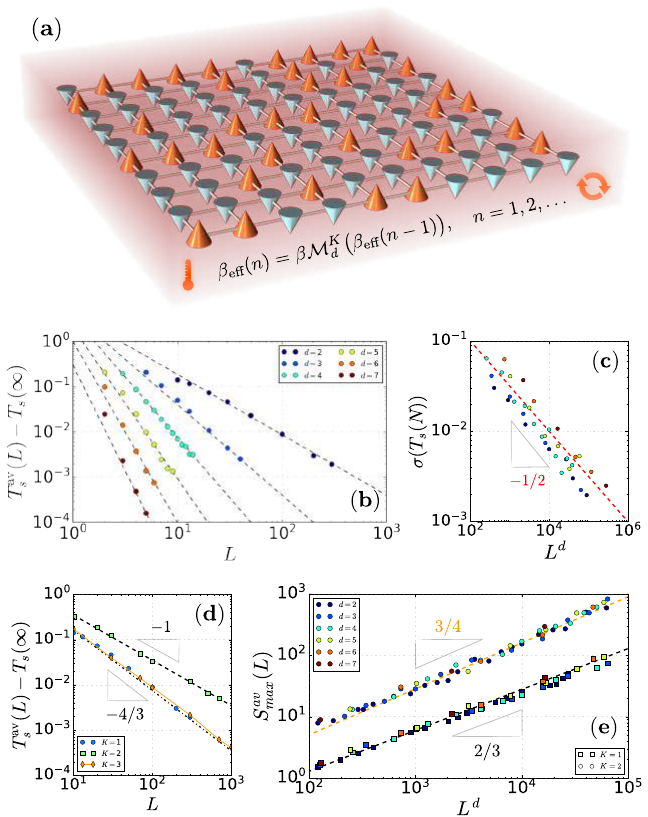}
		\caption{\footnotesize{\textbf{Thermo-adaptive Ising lattices.} 
		(\textbf{a}) Illustration of the model on a square lattice in $d=2$, where the heat-bath temperature, $T\equiv1/\beta$, and the global magnetization, $\mathcal{M}$, form a feedback loop, $\beta_n=\beta\mathcal{M}_{n-1}^K$ with $K=0,1,2\,\dots$, triggering long-range avalanches and mixed-order FP transitions for any $K\geq1$. 
		(\textbf{b}) FSS of the mean critical width in the parity-breaking model, i.e.\ $K=1$, compared to the predicted scaling in Eq.~\eqref{eq:2} with $\nu_d=3/2d$ for $d=2,\dots,7$. 
		(\textbf{c}) Fluctuations of the pseudocritical temperatures, $\sigma(T_s(N))$, satisfy the white-noise scaling, i.e.\ $\propto L^{-d/2}=N^{-1/2}$. 
		(\textbf{d}) Switch between $\phi^3_d$-cascades ($K$ odd) and $\phi^6_d$-cascades ($K$ even) in the correlation length exponent as in Eq.~\eqref{eq:1}. 
		(\textbf{e}) FSS of the mean mass of large finite avalanches sampled during the metastable relaxation slightly above $T_s(L)$ (see text). Results for $K=1$ and $K=2$ follow the expected switching in FSS, Eq.~\eqref{eq:1}. 
		}} \label{fig:3}
\end{figure}

We now focus on the fractal dimension, $\mathrm{D}_d$, of {\em critical} $\phi_d^3$-cascades. Some extra care is needed since, at the transition threshold, one infinite avalanche engulfs the whole system, inheriting its embedding dimension $d$. To resolve this, we study the mass, $s_a^{\mathrm{max}}$, of the largest {\em finite} avalanche formed at $a_s(L)$. Since these are finite-size realizations of spinodal fluctuations, we expect that 
\begin{equation}\label{eq:3}
s_a^{\text{max}}\big(a_s(L)\big)\propto L^{\mathrm{D}_d},
\end{equation}
where $\mathrm{D}_d=2d/3$. Equivalently, in \textsc{Kenna} {\em et al.}\ scheme, the r.h.s.\ of Eq.~\eqref{eq:3} would read $L^{\text{\Koppa}\mathrm{D}}$ with $\text{\Koppa}=d/6$ and $\mathrm{D}=4$. As shown in Fig.~\ref{fig:2}\textbf{d},\,\textbf{e}, finite critical cascades in both processes satisfy the fractal scaling in Eq.~\eqref{eq:3}, while infinite avalanches are compact (Fig.~\ref{fig:2}\textbf{d}, inset). 

\begin{center}
\textsc{Role of symmetry}\vspace*{-0.25cm}
\end{center}
$\,\,\,$ Having clarified hyperscaling and the role of white-noise, we can return to the assumption that long-range cascades physically realize spinodal fluctuations. Based on this, in $\mathrm{R_1}$ we postulated about a second class of $\phi^6_d$-cascades in processes endowed with $\mathrm{Z}_2$ symmetry undergoing mixed-order phase transitions. 

To test this novel scenario, we introduce here a thermo-adaptive Ising model (Fig.~\ref{fig:3}\textbf{a}) where parity is controlled via a thermal coupling. In a nutshell (details in \textsc{Methods}), the dynamics of this model unfolds via a recursive sequence of spin-flip (e.g.\ Glauber) kinetics of a classic Ising lattice in $d$ dimensions, where a global feedback is set after each equilibration stage by the effective temperature $\beta_\mathrm{eff}=\beta f(m_\mathrm{eq})$, where $f(x)=x^K$ is a coupling function of the average magnetization $m_\mathrm{eq}$. For $K>0$, this yields a recursion of temperatures (see Eqs.~\eqref{eq:M1},\,\eqref{eq:M2}) that triggers a long-range propagation of paramagnetic avalanches. It is worth to notice that such a global coupling is natural for disease or opinion spreading, as well as for experimental realizations (see \textsc{Applications}). 
The model undergoes a mixed-order ferro-paramagnetic (\textsc{FP}) transition with spontaneous $\mathds{Z}_2$ symmetry breaking for $K=2n$ with $n\geq1$, and forcedly broken symmetry (i.e.\ $\phi^3$) otherwise (\textsc{Methods}), making it best suited for studying the influence of global symmetries on the critical behavior of long-range avalanches. 

In light of Eq.~\eqref{eq:1}, we expect $T_s^\mathrm{av}(L)-T_s^\infty\propto L^{-1/\nu_d(K)}$ where $\nu_d(K)=3/2d$ if $K=2n+1$ and $\nu_d(K)=2/d$ if $K=2n$, for any integer $n\in\mathds{N}_0$. As a testing ground, we start with $K=1$, whose results on $d$-dimensional lattices are shown in Fig.~\ref{fig:3}\textbf{b}. As expected, critical cascades belong to the $\phi^3_d$ class in all the tested dimensions and the critical width, $\sigma(T_s(L))$, bears the white-noise dominance over critical fluctuations (Fig.~\ref{fig:3}\textbf{c}). Next, we test the role of the $\mathds{Z}_2$-invariance. For simplicity, we study the model for $d=2$ and increase the order, $K$, of its thermo-adaptive coupling. For $K=1,2,3$, $\nu_2$ has to switch between $\nu_2=3/4$ and $\nu_2=1$ as we increase $K$. Results in Fig.~\ref{fig:3}\textbf{d} confirm this alternating behavior, with the model switching from $\phi^3_2$ to $\phi^6_2$ cascades. 

To corroborate the above, we further study the fractal dimension of the largest finite avalanche formed at $T_s(L)$. Identifying the latter calls for a generalized \textsc{Coniglio}-\textsc{Klein} (CK) mapping~\cite{coniglio1980clusters, ray1990nucleation} of our thermo-adaptive Ising model onto some suitable correlated percolation problem, a challenge beyond the scope of this Letter (see \textsc{Discussions}). To circumvent this, we analyze the variations of the total magnetization, $s_t\equiv\mathcal{M}_{t+1}-\mathcal{M}_t$, during the relaxation of the system to the paramagnetic phase slightly above $T_s(L)$. The intuition behind it is that, when deep quenching the system from $\mathcal{M}/N=1$ just above $T_s(L)$, its relaxation slows down close to the jump of the mixed-order transition ---i.e.\ near the saddle-node bifurcation--- leading to a metastable kinetic regime characterized by critical branching~\cite{dong-pre2014, baxter2015critical}. During this long-lived ``plateau'', growing and decaying avalanches of paramagnetic clusters are, themselves, finite-size realizations of spinodal fluctuations whose average mass, we argue, follows the fractal FSS in Eq.~\eqref{eq:3}. We test this method on $K=1$ and $K=2$ thermo-adaptive Ising lattices in $d=2,\dots7$ dimensions by deep quenching the system at $T-T_s(L)=\delta$ for suitable values of $\delta\leq10^{-3}$. Figure \ref{fig:3}\textbf{e} highlights our results, plotted against the lattices' volume $N=L^d$ to ease the exposition (see also Fig.~\ref{fig:S1}). As visible, the mass of finite avalanches nicely follows the switching of the fractal dimensions predicted in Eq.~\eqref{eq:1} between $\phi^3_d$ and $\phi^6_d$-cascades. 

Furthermore, in Figs.~\ref{fig:S2},\,\ref{fig:S3} in the \textsc{Supplementary Material} (SM) we confirm that the distribution of critical avalanches sampled during the metastable plateau regime satisfies the power-law scaling $\tau(s)\propto s^{-\tau}$ with the Fisher's exponents $\tau=1+d/\mathrm{D}_d=\{5/2,7/3\}$ expected, respectively, for $\phi^3_d$ and $\phi^6_d$ cascades.

\begin{center}
\textsc{Applications}\vspace*{-0.25cm}
\end{center} 
$\,\,\,$ Among many potential applications (\textsc{Discussions}), we highlight here disease spreading and electric runaway. 

\begin{figure}[t]\vspace*{-0.25cm}
	\includegraphics[width=\linewidth]{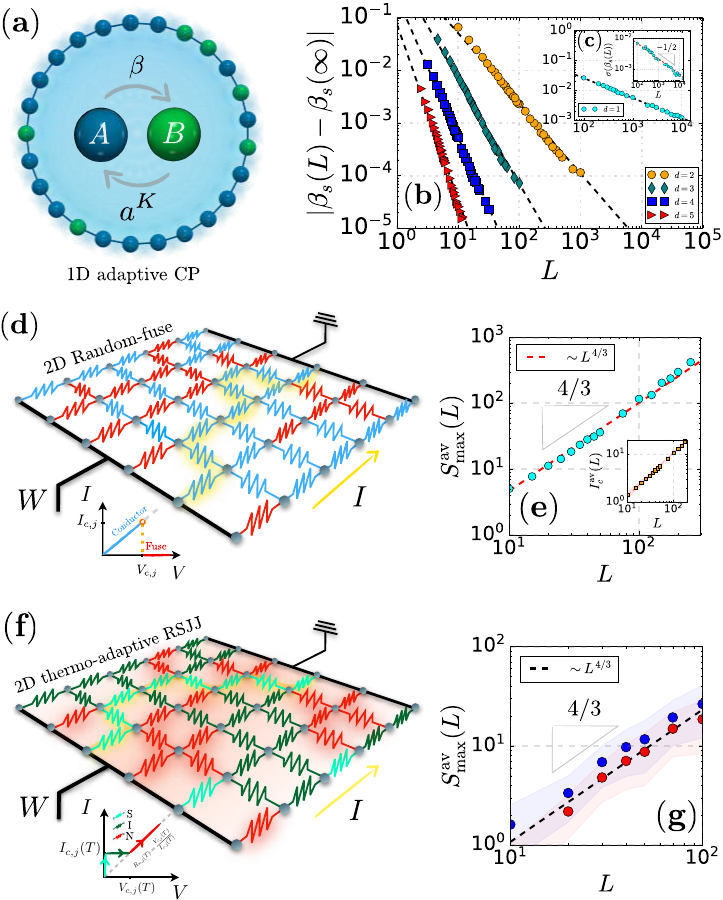}
		\caption{\footnotesize{ \textbf{Applications}. 
		(\textbf{a}) Adaptive contact process (aCP) undergoing mixed-order A-to-B transitions (Figs.~\ref{fig:S4}\textbf{a}--\textbf{c}, SM) for any $K\geq2$ and $d\geq1$. 
		(\textbf{b}, \textbf{c}) FSS of the mean critical width for $K=2$ on lattices for $d=1,\dots,5$. 
		(Inset, \textbf{c}) Fluctuations of the rates $\beta_s(L)$ in $d=1$ confirming the scaling $\sigma\propto L^{-1/2}$. 
		(\textbf{d}) Random fuse (RF) model in $d=2$; (inset) voltage-current characteristic of the bonds. 
		(\textbf{e}) FSS of the critical avalanches sampled during the relaxation from the fully conductive state, $G/N=1$, to the fused state, $G/N\to0$, slightly above $I_c(L)$. 
		(Inset) Critical currents safisfy the FSS relation~\cite{zapperi1999avalanches} $I_c(L)\propto L\ln L$ (\textsc{Methods}). 
		(\textbf{f}) Thermo-resistive network (see text) modeling interdependent superconductors~\cite{bonamassa2023interdependent} ; (inset) voltage-current characteristic of the bonds. 
		(\textbf{g}) FSS of the largest finite avalanches sampled after quenching the circuit to the fully metal phase slightly above $I_c(L)$; metallic (red markers) and superconducting (blue markers) avalanches satisfy the fractal scaling in Eq.~\eqref{eq:3} with dimension $\mathrm{D}_2=4/3$. 
		}} \label{fig:4}
\end{figure}

Long-range cascades are, in fact, relevant in opinion formation and contagion dynamics~\cite{iacopini2019simplicial}, reason why we test our classification on an adaptive $2$-state ($A$ and $B$, Fig.~\ref{fig:4}\textbf{a}) contact process (CP). In this model (see \textsc{Methods}), the state of the $i$-th node changes from $A\to B$ with rate $\beta\in[0,1]$ and from $B\to A$ with global adaptive rate $\mathcal{A}^K\in[0,1]$, where $K\in\mathds{N}_0$ and $\mathcal{A}$ is the fraction of $A$-nodes in the network, if at least one of its neighbors has state $A$. This CP could describe e.g.\ SIS or failure-recovery~\cite{majdandzic2014spontaneous, bottcher2017critical, danziger2019dynamic} processes, whose reinfection rate weakens adaptively with the global infectivity. The model undergoes mixed-order $A$-to-$B$ transitions for $K\geq1$ and $d\geq1$ (Figs.~\ref{fig:S4}\textbf{a}--\textbf{c}, SM), and it is not $\mathds{Z}_2$ invariant; thus, we expect it to exhibit $\phi^3_d$ cascades. As shown in Figs.~\ref{fig:4}\textbf{b},\textbf{c}, simulations on lattices with $d=1,\dots,5$ confirm Eq.~\eqref{eq:2} and the white-noise dominance (Fig.~\ref{fig:4}\textbf{c}, inset). Figure \ref{fig:S4}\textbf{e},\textbf{f} in the SM further confirm the fractal dimension $\mathrm{D}_d=2d/3$ and the Fisher's exponent $\tau=5/2$. 

We conclude by studying two experimentally-driven models, one for fractures~\cite{pradhan2010failure, hansen2015fiber} and the other one for interdependent superconductors~\cite{bonamassa2023interdependent}, where long-range cascades form in response to emergent (i.e.\ not superimposed, as in the models above) long-range feedback. For the former case, we consider the random-fuse (RF) model~\cite{de1985random, duxbury1986size, kahng1988electrical} (Fig.~\ref{fig:4}\textbf{d}), where the electric runaway due to the burning of bonds during the redistribution of the current yields the formation of avalanches satisfying spinodal scaling~\cite{zapperi1997first, zapperi1999avalanches}. For the second case, instead, we adopt the thermo-resistively-shunted Josephson-junction model (tRSSJ, \textsc{Methods}) introduced in Ref.~\cite{bonamassa2023interdependent} (Figs.~\ref{fig:4}\textbf{f}), where mixed-order transitions emerge due to a thermo-electric feedback. The micro-irreversible nature of both processes predicts, here again, long-range $\phi^3_d$ cascades. We test this in $d=2$ by solving iteratively the Kirchhoff equations (\textsc{Methods}) for increasing values of the bias current, $I_b$, under the characteristic voltage-current profiles sketched, respectively, in Figs.~\ref{fig:4}\textbf{d},\,\textbf{f}. Like in the thermo-adaptive Ising model, we study the FSS of the mass of avalanches, Eq.~\eqref{eq:3}, sampled slightly above the critical current, $I_c(L)$, during the metastable relaxation of the circuit towards the metal/fused phase. As shown in Figs.~\ref{fig:4}\textbf{e},\textbf{g}, despite the relatively small sizes, both models exhibit the fractal dimension $\mathrm{D}_2=4/3$ expected for $\phi^3_2$ long-range cascades (see Fig.~\ref{fig:2}\textbf{d},\textbf{e}). 

\begin{center}
\textsc{Discussions}\vspace*{-0.25cm}
\end{center}
$\,\,\,$ We showed that long-range cascades at mixed-order transitions with exponent $\beta=1/2$ realize mean-field spinodal fluctuations. Their critical properties are {\em hybrid}, i.e.\ they are spinodal (mean-field) with upper critical dimension $d_c=d$ or, equivalently, with the FSS-exponent~\cite{berche2012hyperscaling, kenna2014fisher, kenna2017universal, berche2022phase} $\text{\Koppa}=d/d_c$, {\em for any} $d\gtrless d_c$. We demonstrate the existence of two classes, denoted $\phi^3_d$ and $\phi^6_d$, characterized by the $\mathds{Z}_2$ invariance of the cascading process, and showed how symmetry influences the geometry and evolution of avalanches. In particular, $\phi^6_d$-cascades have larger mass, longer correlation length and longer lifetime, $\tau(L)\propto L^{1/2\nu_d}$, compared to $\phi^3_d$ ones. 

We have supported these critical features using extensive simulations over paradigmatic models for structural, thermal and dynamic long-range cascades, though we expect our classification to encompass a much broader range of processes undergoing mixed-order or explosive transitions~\cite{boccaletti2016explosive, d2019explosive, kuehn2021universal}, particularly in systems endowed with higher-order dynamics~\cite{iacopini2019simplicial, battiston2021physics, zhang2023higher, ferraz2023multistability}. Given the role of disorder and global symmetries, we do not rule out the existence of other classes of long-range cascades, whose exponents might differ from those studied here. 

Our results push the study of first-order transitions~\cite{binder1987theory} up to the limit of metastability (i.e.\ the spinodal point), offering the opportunity to explore universal properties lurking fundamental phenomena like, e.g.\ the emergence of computational complexity or ergodicity breaking in glassy and disordered media~\cite{bonamassa2021interdependent}. We expect the hybrid critical phenomena of $\phi^3_d$ long-range cascades to characterize the clustering transition in random $p$-\textsc{xorsat}~\cite{leone2001phase,mezard2003two}, a paradigmatic class of combinatorial optimization problems, whilst $\phi^6_d$ long-range cascades may underlie the dielectric properties of ferroelectric materials~\cite{devonshire1954theory, hoffmann2019unveiling}. We also anticipate that the alternating behavior unveiled here in the $K$-thermo-adaptive Ising lattice characterizes also the dynamic phase transition~\cite{leone2001phase,mezard2003two} of ferromagnetic $p$-spin models on both random hypergraphs (\textsc{Methods}) and on $d$R-hypergraphs with $p=K+2$ and for any $d\geq2$. 
Furthermore, our results for the random-fuse model (Fig.~\ref{fig:4}\textbf{d}) and its connection with the so-called fiber-bundle model for fractures~\cite{pradhan2010failure, hansen2015fiber}, foster perspectives of potential applications of the above in the study of earthquakes and material stability~\cite{rundle2003statistical}. 

In conclusion, our results challenge the traditional view of spinodals as mere mean-field constructs, revealing their physical manifestation through long-range cascades at mixed-order transitions for any dimension, $d$, of the underlying network. This indicates that critical phenomena above their upper critical dimension~\cite{berche2022phase} can manifest their signatures through long-range cascades---e.g.\ mean-field percolation~\cite{kenna2017universal} via $\phi^3_d$ avalanches---even in low dimensions, tracing avenues of theoretical and computational research amenable of experimental validation. 




\FloatBarrier
\bibliographystyle{unsrt}
\bibliography{D_spino.bib}

\begin{thebibliography}{10}

\bibitem{motter2017unfolding}
A.~Motter and Y.~Yang.
\newblock The unfolding and control of network cascades.
\newblock {\em Physics Today}, 70(1):32--39, 2017.

\bibitem{deangelis2012positive}
D.~L. DeAngelis, W.~M. Post, and C.~C. Travis.
\newblock {\em Positive feedback in natural systems}, volume~15.
\newblock Springer Science \& Business Media, 2012.

\bibitem{dodds2004universal}
P.~S. Dodds and D.~J. Watts.
\newblock Universal behavior in a generalized model of contagion.
\newblock {\em Phys. Rev. Lett.}, 92(21):218701, 2004.

\bibitem{haldane2011systemic}
A.~G. Haldane and R.~M. May.
\newblock Systemic risk in banking ecosystems.
\newblock {\em Nature}, 469(7330):351, 2011.

\bibitem{barnosky2012approaching}
A.~D. Barnosky et~al.
\newblock Approaching a state shift in earth’s biosphere.
\newblock {\em Nature}, 486(7401):52--58, 2012.

\bibitem{brummitt2015coupled}
C.~D. Brummitt, G.~Barnett, and R.~M. D'Souza.
\newblock Coupled catastrophes: sudden shifts cascade and hop among
  interdependent systems.
\newblock {\em Journal of The Royal Society Interface}, 12(112):20150712, 2015.

\bibitem{majdandzic2016multiple}
A.~Majdandzic, L.~A. Braunstein, C.~Curme, I.~Vodenska, S.~Levy-Carciente,
  S.~H. Stanley, and S.~Havlin.
\newblock Multiple tipping points and optimal repairing in interacting
  networks.
\newblock {\em Nature communications}, 7(1):10850, 2016.

\bibitem{schafer2018dynamically}
B.~Sch{\"a}fer, D.~Witthaut, M.~Timme, and V.~Latora.
\newblock Dynamically induced cascading failures in power grids.
\newblock {\em Nature communications}, 9(1):1975, 2018.

\bibitem{rocha2018cascading}
J.~C. Rocha, G.~Peterson, O.~Bodin, and S.~A. Levin.
\newblock Cascading regime shifts within and across scales.
\newblock {\em Science}, 362(6421):1379--1383, 2018.

\bibitem{artime2020abrupt}
O.~Artime and M.~De Domenico.
\newblock Abrupt transition due to non-local cascade propagation in multiplex
  systems.
\newblock {\em New Journal of Physics}, 22(9):093035, 2020.

\bibitem{scheffer2020critical}
M.~Scheffer.
\newblock {\em Critical transitions in nature and society}, volume~16.
\newblock Princeton University Press, 2020.

\bibitem{boccaletti2016explosive}
S.~Boccaletti, J.~A. Almendral, S.~Guan, I.~Leyva, Z.~Liu,
  I.~Sendi{\~n}a-Nadal, Z.~Wang, and Y.~Zou.
\newblock Explosive transitions in complex networks’ structure and dynamics:
  Percolation and synchronization.
\newblock {\em Physics Reports}, 660:1--94, 2016.

\bibitem{d2019explosive}
R.~M. D'Souza, J.~G{\'o}mez-Gardenes, J.~Nagler, and A.~Arena.
\newblock Explosive phenomena in complex networks.
\newblock {\em Advances in Physics}, 68(3):123--223, 2019.

\bibitem{kuehn2021universal}
C.~Kuehn and C.~Bick.
\newblock A universal route to explosive phenomena.
\newblock {\em Science advances}, 7(16):eabe3824, 2021.

\bibitem{gross2023dynamics}
B.~Gross, I.~Bonamassa, and S.~Havlin.
\newblock Dynamics of cascades in spatial interdependent networks.
\newblock {\em Chaos: An Interdisciplinary Journal of Nonlinear Science},
  33(10), 2023.

\bibitem{buldyrev-nature2010}
S.~V. Buldyrev, R.~Parshani, G.~Paul, H.~E. Stanley, and S.~Havlin.
\newblock {Catastrophic cascade of failures in interdependent networks}.
\newblock {\em Nature}, 464(7291):1025--1028, 2010.

\bibitem{bashan2013extreme}
A.~Bashan, Y.~Berezin, S.~V. Buldyrev, and S.~Havlin.
\newblock {The extreme vulnerability of interdependent spatially embedded
  networks}.
\newblock {\em Nature Physics}, 9:667--672, Aug 2013.

\bibitem{iacopini2019simplicial}
I.~Iacopini, G.~Petri, A.~Barrat, and V.~Latora.
\newblock Simplicial models of social contagion.
\newblock {\em Nature communications}, 10(1):2485, 2019.

\bibitem{battiston2021physics}
F.~Battiston, E.~Amico, A.~Barrat, G.~Bianconi, G.~Ferraz de~Arruda,
  B.~Franceschiello, I.~Iacopini, S.~K{\'e}fi, V.~Latora, Y.~Moreno, et~al.
\newblock The physics of higher-order interactions in complex systems.
\newblock {\em Nature Physics}, 17(10):1093--1098, 2021.

\bibitem{zhang2023higher}
Y.~Zhang, M.~Lucas, and F.~Battiston.
\newblock Higher-order interactions shape collective dynamics differently in
  hypergraphs and simplicial complexes.
\newblock {\em Nature Communications}, 14(1):1605, 2023.

\bibitem{ferraz2023multistability}
G.~Ferraz de~Arruda, G.~Petri, P.~M. Rodriguez, and Y.~Moreno.
\newblock Multistability, intermittency, and hybrid transitions in social
  contagion models on hypergraphs.
\newblock {\em Nature Communications}, 14(1):1375, 2023.

\bibitem{zeng2020multiple}
G.~Zeng, J.~Gao, L.~Shekhtman, S.~Guo, W.~Lv, J.~Wu, H.~Liu, O.~Levy, D.~Li,
  Z.~Gao, et~al.
\newblock Multiple metastable network states in urban traffic.
\newblock {\em Proceedings of the National Academy of Sciences},
  117(30):17528--17534, 2020.

\bibitem{motter2002cascade}
A.~E. Motter and Y.-C. Lai.
\newblock Cascade-based attacks on complex networks.
\newblock {\em Physical Review E}, 66(6):065102, 2002.

\bibitem{hoffmann2014suppressing}
H.~Hoffmann and D.~W. Payton.
\newblock Suppressing cascades in a self-organized-critical model with
  non-contiguous spread of failures.
\newblock {\em Chaos, Solitons \& Fractals}, 67:87--93, 2014.

\bibitem{zhao2016spatio}
J.~Zhao, D.~Li, H.~Sanhedrai, R.~Cohen, and S.~Havlin.
\newblock Spatio-temporal propagation of cascading overload failures in
  spatially embedded networks.
\newblock {\em Nature communications}, 7(1):10094, 2016.

\bibitem{henkes2005jamming}
S.~Henkes and B.~Chakraborty.
\newblock Jamming as a critical phenomenon: A field theory of zero-temperature
  grain packings.
\newblock {\em Phys. Rev. Lett.}, 95(19):198002, 2005.

\bibitem{silbert2005vibrations}
L.~E. Silbert, A.~J. Liu, and S.~R. Nagel.
\newblock Vibrations and diverging length scales near the unjamming transition.
\newblock {\em Physical Review Letters}, 95(9):098301, 2005.

\bibitem{schwarz2006onset}
J.~M. Schwarz, A.~J. Liu, and L.~Q. Chayes.
\newblock The onset of jamming as the sudden emergence of an infinite k-core
  cluster.
\newblock {\em Europhysics Letters}, 73(4):560, 2006.

\bibitem{ferguson1999spinodals}
C.~D. Ferguson, W.~Klein, and J.~B. Rundle.
\newblock Spinodals, scaling, and ergodicity in a threshold model with
  long-range stress transfer.
\newblock {\em Physical Review E}, 60(2):1359, 1999.

\bibitem{petri1994experimental}
A.~Petri, G.~Paparo, A.~Vespignani, A.~Alippi, and M.~Costantini.
\newblock Experimental evidence for critical dynamics in microfracturing
  processes.
\newblock {\em Phys. Rev. Lett.}, 73(25):3423, 1994.

\bibitem{moreno2000fracture}
Y.~Moreno, J.~B. Gomez, and A.~F. Pacheco.
\newblock Fracture and second-order phase transitions.
\newblock {\em Phys. Rev. Lett.}, 85(14):2865, 2000.

\bibitem{rundle2003statistical}
J.~B. Rundle, D.~L. Turcotte, R.~Shcherbakov, W.~Klein, and C.~Sammis.
\newblock Statistical physics approach to understanding the multiscale dynamics
  of earthquake fault systems.
\newblock {\em Reviews of Geophysics}, 41(4), 2003.

\bibitem{heermann1982spinodals}
D.~W. Heermann, W.~Klein, and D.~Stauffer.
\newblock Spinodals in a long-range interaction system.
\newblock {\em Phys. Rev. Lett.}, 49(17):1262, 1982.

\bibitem{unger1984nucleation}
C.~Unger and W.~Klein.
\newblock Nucleation theory near the classical spinodal.
\newblock {\em Physical Review B}, 29(5):2698, 1984.

\bibitem{monette1994spinodal}
L~Monette.
\newblock Spinodal nucleation.
\newblock {\em International Journal of Modern Physics B}, 8(11n12):1417--1527,
  1994.

\bibitem{klein2007structure}
W.~Klein, H.~Gould, N.~Gulbahce, J.~B. Rundle, and K.~Tiampo.
\newblock Structure of fluctuations near mean-field critical points and
  spinodals and its implication for physical processes.
\newblock {\em Physical Review E}, 75(3):031114, 2007.

\bibitem{gross2022fractal}
B.~Gross, I.~Bonamassa, and S.~Havlin.
\newblock Fractal fluctuations at mixed-order transitions in interdependent
  networks.
\newblock {\em Phys. Rev. Lett.}, 129(26):268301, 2022.

\bibitem{stauffer_book}
D~Stauffer.
\newblock {\em Introduction to Percolation Theory}.
\newblock London, Francis \& Taylor, 1985.

\bibitem{binder1987theory}
K.~Binder.
\newblock Theory of first-order phase transitions.
\newblock {\em Reports on progress in physics}, 50(7):783, 1987.

\bibitem{aizenman1997number}
Michael Aizenman.
\newblock On the number of incipient spanning clusters.
\newblock {\em Nuclear Physics B}, 485(3):551--582, 1997.

\bibitem{coniglio1985shapes}
A.~Coniglio.
\newblock Shapes, surfaces, and interfaces in percolation clusters.
\newblock In {\em Physics of Finely Divided Matter: Proceedings of the Winter
  School, Les Houches, France, March 25--April 5, 1985}, pages 84--101.
  Springer, 1985.

\bibitem{binder1985finite}
K.~Binder, M.~Nauenberg, V.~Privman, and A.~P. Young.
\newblock Finite-size tests of hyperscaling.
\newblock {\em Physical Review B}, 31(3):1498, 1985.

\bibitem{binder1985critical}
K.~Binder.
\newblock Critical properties and finite-size effects of the five-dimensional
  ising model.
\newblock {\em Zeitschrift f{\"u}r Physik B Condensed Matter}, 61:13--23, 1985.

\bibitem{luijten1996finite}
E.~Luijten and H.~W.~J. Bl{\"o}te.
\newblock Finite-size scaling and universality above the upper critical
  dimensionality.
\newblock {\em Physical review letters}, 76(10):1557, 1996.

\bibitem{parisi1996scaling}
G.~Parisi and J.~J. Ruiz-Lorenzo.
\newblock Scaling above the upper critical dimension in ising models.
\newblock {\em Physical Review B}, 54(6):R3698, 1996.

\bibitem{berche2012hyperscaling}
B.~Berche, R.~Kenna, and J.-C. Walter.
\newblock Hyperscaling above the upper critical dimension.
\newblock {\em Nuclear Physics B}, 865(1):115--132, 2012.

\bibitem{kenna2014fisher}
R.~Kenna and B.~Berche.
\newblock Fisher's scaling relation above the upper critical dimension.
\newblock {\em Europhysics Letters}, 105(2):26005, 2014.

\bibitem{kenna2017universal}
R.~Kenna and B.~Berche.
\newblock Universal finite-size scaling for percolation theory in high
  dimensions.
\newblock {\em Journal of Physics A: Mathematical and Theoretical},
  50(23):235001, 2017.

\bibitem{berche2022phase}
B.~Berche, T.~Ellis, Y.~Holovatch, and R.~Kenna.
\newblock Phase transitions above the upper critical dimension.
\newblock {\em SciPost Physics Lecture Notes}, page 060, 2022.

\bibitem{fisher1967theory}
M.~E. Fisher.
\newblock The theory of equilibrium critical phenomena.
\newblock {\em Reports on progress in physics}, 30(2):615, 1967.

\bibitem{kirkpatrick2015exponent}
T.~R. Kirkpatrick and D.~Belitz.
\newblock Exponent relations at quantum phase transitions with applications to
  metallic quantum ferromagnets.
\newblock {\em Physical Review B}, 91(21):214407, 2015.

\bibitem{chayes1986finite}
J.~T. Chayes, L.~Chayes, D.~S. Fisher, and T.~Spencer.
\newblock Finite-size scaling and correlation lengths for disordered systems.
\newblock {\em Phys. Rev. Lett.}, 57(24):2999, 1986.

\bibitem{lee2016hybrid}
D.~Lee, S.~Choi, M.~Stippinger, J.~Kert{\'e}sz, and B.~Kahng.
\newblock Hybrid phase transition into an absorbing state: Percolation and
  avalanches.
\newblock {\em Physical Review E}, 93(4):042109, 2016.

\bibitem{bonamassa2021interdependent}
I.~Bonamassa, B.~Gross, and S.~Havlin.
\newblock Interdependent couplings map to thermal, higher-order interactions.
\newblock {\em arXiv preprint arXiv:2110.08907}, 2021.

\bibitem{leone2001phase}
M.~Leone, F.~Ricci-Tersenghi, and R.~Zecchina.
\newblock Phase coexistence and finite-size scaling in random combinatorial
  problems.
\newblock {\em Journal of Physics A: Mathematical and General}, 34(22):4615,
  2001.

\bibitem{mezard2003two}
M.~M{\'e}zard, F.~Ricci-Tersenghi, and R.~Zecchina.
\newblock Two solutions to diluted p-spin models and xorsat problems.
\newblock {\em Journal of Statistical Physics}, 111:505--533, 2003.

\bibitem{coniglio1980clusters}
A.~Coniglio and W.~Klein.
\newblock Clusters and ising critical droplets: a renormalisation group
  approach.
\newblock {\em J. Phys. A: Math. and Gen.}, 13(8):2775, 1980.

\bibitem{ray1990nucleation}
T.~S. Ray and W.~Klein.
\newblock Nucleation near the spinodal in long-range ising models.
\newblock {\em Journal of statistical physics}, 61:891--902, 1990.

\bibitem{dong-pre2014}
D.~Zhou, A.~Bashan, R.~Cohen, Y.~Berezin, N.~Shnerb, and S.~Havlin.
\newblock Simultaneous first- and second-order percolation transitions in
  interdependent networks.
\newblock {\em Phys. Rev. E}, 90:012803, Jul 2014.

\bibitem{baxter2015critical}
G.~J. Baxter, S.~N. Dorogovtsev, K.-E. Lee, J.~F.~F. Mendes, and A.~V. Goltsev.
\newblock Critical dynamics of the k-core pruning process.
\newblock {\em Phys. Rev. X}, 5(3):031017, 2015.

\bibitem{majdandzic2014spontaneous}
A.~Majdandzic, B.~Podobnik, S.~V. Buldyrev, D.~Y. Kenett, S.~Havlin, and E.~H.
  Stanley.
\newblock Spontaneous recovery in dynamical networks.
\newblock {\em Nature Physics}, 10(1):34--38, 2014.

\bibitem{bottcher2017critical}
L.~B{\"o}ttcher, J.~Nagler, and H.~J. Herrmann.
\newblock Critical behaviors in contagion dynamics.
\newblock {\em Phys. Rev. Lett.}, 118(8):088301, 2017.

\bibitem{danziger2019dynamic}
M.~M. Danziger, I.~Bonamassa, S.~Boccaletti, and S.~Havlin.
\newblock Dynamic interdependence and competition in multilayer networks.
\newblock {\em Nature Physics}, 15(2):178, 2019.

\bibitem{zapperi1999avalanches}
S.~Zapperi, P.~Ray, H.~E. Stanley, and A.~Vespignani.
\newblock Avalanches in breakdown and fracture processes.
\newblock {\em Physical Review E}, 59(5):5049, 1999.

\bibitem{bonamassa2023interdependent}
I.~Bonamassa, B.~Gross, M.~Laav, I.~Volotsenko, A.~Frydman, and S.~Havlin.
\newblock Interdependent superconducting networks.
\newblock {\em Nature Physics}, pages 1--8, 2023.

\bibitem{pradhan2010failure}
S.~Pradhan, A.~Hansen, and B.~K. Chakrabarti.
\newblock Failure processes in elastic fiber bundles.
\newblock {\em Reviews of modern physics}, 82(1):499, 2010.

\bibitem{hansen2015fiber}
A.~Hansen, P.~C. Hemmer, and S.~Pradhan.
\newblock {\em The fiber bundle model: modeling failure in materials}.
\newblock John Wiley \& Sons, 2015.

\bibitem{de1985random}
L.~de~Arcangelis, S.~Redner, and H.~J. Herrmann.
\newblock A random fuse model for breaking processes.
\newblock {\em Journal de Physique Lettres}, 46(13):585--590, 1985.

\bibitem{duxbury1986size}
P.~M. Duxbury, P.~D. Beale, and P.~L. Leath.
\newblock Size effects of electrical breakdown in quenched random media.
\newblock {\em Physical review letters}, 57(8):1052, 1986.

\bibitem{kahng1988electrical}
B.~Kahng, G.~G. Batrouni, S.~Redner, L.~de~Arcangelis, and H.~J. Herrmann.
\newblock Electrical breakdown in a fuse network with random, continuously
  distributed breaking strengths.
\newblock {\em Physical Review B}, 37(13):7625, 1988.

\bibitem{zapperi1997first}
S.~Zapperi, P.~Ray, H.~E. Stanley, and A.~Vespignani.
\newblock First-order transition in the breakdown of disordered media.
\newblock {\em Phys. Rev. Lett.}, 78(8):1408, 1997.

\bibitem{devonshire1954theory}
A.~F. Devonshire.
\newblock Theory of ferroelectrics.
\newblock {\em Advances in physics}, 3(10):85--130, 1954.

\bibitem{hoffmann2019unveiling}
M.~Hoffmann {\em et al.}
\newblock Unveiling the double-well energy landscape in a ferroelectric layer.
\newblock {\em Nature}, 565(7740):464--467, 2019.

\bibitem{baxter2011onsager}
R.~J. Baxter.
\newblock Onsager and $k$aufman’s calculation of the spontaneous
  magnetization of the ising model.
\newblock {\em Journal of Statistical Physics}, 145:518--548, 2011.

\bibitem{baxter2012onsager}
R.~J. Baxter.
\newblock Onsager and $k$aufman’s calculation of the spontaneous
  magnetization of the ising model: Ii.
\newblock {\em Journal of Statistical Physics}, 149:1164--1167, 2012.

\bibitem{pastor2015epidemic}
R.~Pastor-Satorras, C.~Castellano, P.~Van Mieghem, and A.~Vespignani.
\newblock Epidemic processes in complex networks.
\newblock {\em Reviews of Modern Physics}, 87(3):925, 2015.

\bibitem{zapperi1995self}
Stefano Zapperi, Kent~B{\ae}kgaard Lauritsen, and H~Eugene Stanley.
\newblock Self-organized branching processes: mean-field theory for avalanches.
\newblock {\em Physical review letters}, 75(22):4071, 1995.

\bibitem{zapperi1997plasticity}
S.~Zapperi, A.~Vespignani, and H.~E. Stanley.
\newblock Plasticity and avalanche behaviour in microfracturing phenomena.
\newblock {\em Nature}, 388(6643):658--660, 1997.

\bibitem{fisher1982scaling}
Michael~E Fisher and A~Nihat Berker.
\newblock Scaling for first-order phase transitions in thermodynamic and finite
  systems.
\newblock {\em Physical Review B}, 26(5):2507, 1982.

\bibitem{gross1985mean}
D.~J. Gross, I.~Kanter, and H.~Sompolinsky.
\newblock Mean-field theory of the potts glass.
\newblock {\em Physical review letters}, 55(3):304, 1985.

\bibitem{kirkpatrick1987p}
T.~R. Kirkpatrick and D.~Thirumalai.
\newblock p-spin-interaction spin-glass models: Connections with the structural
  glass problem.
\newblock {\em Physical Review B}, 36(10):5388, 1987.

\bibitem{kirkpatrick2015colloquium}
TR~Kirkpatrick and D~Thirumalai.
\newblock Colloquium: Random first order transition theory concepts in biology
  and physics.
\newblock {\em Reviews of Modern Physics}, 87(1):183, 2015.

\bibitem{kadanoff1967static}
L.~P. Kadanoff, W.~G{\"o}tze, D.~Hamblen, R.~Hecht, E.~A.~S. Lewis, V.~V.
  Palciauskas, M.~Rayl, J.~Swift, D.~Aspnes, and J.~Kane.
\newblock Static phenomena near critical points: theory and experiment.
\newblock {\em Reviews of Modern Physics}, 39(2):395, 1967.

\bibitem{hankey1971alternate}
A.~Hankey and H.~E. Stanley.
\newblock An alternate formulation of the static scaling hypothesis.
\newblock {\em International Journal of Quantum Chemistry}, 5(S5):593--604,
  1971.

\bibitem{piscitelli2021jamming}
A.~Piscitelli, A.~Coniglio, A.~Fierro, and M.~P. Ciamarra.
\newblock Jamming as a random first-order percolation transition.
\newblock {\em Physica A: Statistical Mechanics and its Applications},
  569:125796, 2021.

\bibitem{huang2018critical}
W.~Huang, P.~Hou, J.~Wang, R.~M. Ziff, and Y.~Deng.
\newblock Critical percolation clusters in seven dimensions and on a complete
  graph.
\newblock {\em Physical Review E}, 97(2):022107, 2018.

\bibitem{gross2023microscopic}
B.~Gross, I.~Bonamassa, and S.~Havlin.
\newblock Microscopic intervention yields abrupt transition in interdependent
  magnetic networks.
\newblock {\em arXiv preprint arXiv:2306.05573}, 2023.

\end{thebibliography}

\renewcommand\thesection{M\arabic{section}}
\renewcommand\thesubsection{M\arabic{section}.\arabic{subsection}}
\setcounter{section}{0}
\setcounter{equation}{0}
\setcounter{figure}{0}
\renewcommand{\theequation}{M\arabic{equation}}
\renewcommand{\thefigure}{M\arabic{figure}}

{\small
\section*{Methods}
$\mathrm{M1})$ \textsc{Thermo-adaptive Ising model}. Consider an Ising lattice in $d$-dimensions and let $\boldsymbol{\sigma}_{\mathrm{eq}}(\beta)\in\mathds{Z}_2^N$ be a spin configuration reached by the system at the inverse heat-bath temperature $\beta=1/T$ (Boltzmann units are intended). The thermalization of our thermo-adaptive Ising model evolves by a recursive sequence of macroscopic equilibria and temperature updates until a global steady state is reached. In details, we initiate the system in the fully ferromagnetic state (i.e.\ $\sigma_i=1$ for every $i\leq N$) at the inverse heat-bath temperature $\beta$ and we let it to equilibrate to some spin configuration $\boldsymbol{\sigma}_\mathrm{eq}(\beta)$. We then update $\beta$ via the thermalized magnetization density $m_\mathrm{eq}(\beta)=\frac{1}{N}\sum_{i\leq N}\langle \sigma_i\rangle_{\beta}$, where $\langle\,(\cdots)\rangle_\beta$ is a thermal average, by means of the thermal coupling $\beta_{\mathrm{eff}}\equiv \beta f(m_\mathrm{eq})$, where $f(x)$ is some function of the average magnetization. We then let the Ising model to reach a new equilibrium at the effective temperature $\beta_\mathrm{eff}$. By iterating this scheme, we generate a recursive sequence of adaptive global temperatures
\begin{equation}\label{eq:M1}
\beta\xmapsto{m_\mathrm{eq}(\beta)}
\beta_\mathrm{eff}^{(1)}\xmapsto{m_\mathrm{eq}\left(\beta_\mathrm{eff}^{(1)}\right)} 
\beta_\mathrm{eff}^{(2)}\xmapsto{m_\mathrm{eq}\left(\beta_\mathrm{eff}^{(2)}\right)} 
\cdots,\vspace*{-0.1cm}
\end{equation}
whose evolution underlies a thermomagnetic feedback that can trigger the propagation of paramagnetic avalanches, depending on the coupling function $f$. Since we are interested in positive feedback, we adopt the function $f(x)=x^K$, i.e.\
\begin{equation}\label{eq:M2}
\beta_\mathrm{eff}(n)\equiv \beta m_\mathrm{eq}^K(n-1),\quad K=1,2,3,\dots,
\end{equation}
\noindent 
where $n=2,3,\dots$ denote the thermalization steps needed before reaching a steady state. The power $x^K$ in Eq.~\eqref{eq:M2} controls the parity of the model. This can be checked e.g.\ by implementing Eqs.~\eqref{eq:M1},\,\eqref{eq:M2} for the Glauber dynamics on a complete graph, whose steady state solution yields the self-consistent equation $m=\mathrm{tanh}(\beta m^{K+1})$; a more detailed study, leading to similar results, can be found in Ref.~\cite{bonamassa2021interdependent}. Besides the trivial $K=0$ case (i.e.\ the Curie-Weiss model), positive values of $K$ produce thermo-adaptive Ising models whose $\mathds{Z}_2$ symmetry is broken if $K=2n+1$ and preserved if instead $K=2n$, for every $n\geq0$. As shown in Ref.~\cite{bonamassa2021interdependent}, this switching can be also controlled by thermally coupling multiple layers; in fact, the $K$-thermo-adaptive Ising model for $K\geq1$ is equivalent to $K+1$ thermally dependent Ising layers where Eq.~\eqref{eq:M2} involves, in a suitable way, the global magnetization of other layers. In this case, the $\mathds{Z}_2$ invariance is governed by the number of coupled layers. \vspace*{+0.25cm}

$\mathrm{M2})$ \textsc{Solving thermo-adaptive spin models}. Given the novelty of the thermo-adaptive Ising model, we provide a few details about its analytical solution. The recursion in Eq.~\eqref{eq:M1} uses, at each thermalization step, the single-layer behavior for the effective temperature $\beta_\mathrm{eff}$; hence, we can solve the kinetics of the model by iterating the solution (if known) of the isolated Ising lattice. Assuming that such a solution is of the form $m_\mathrm{eq}(d,\beta)=\mathcal{X}_d(\beta)$, we can rewrite Eq.~\eqref{eq:M2} as
\begin{equation}\label{eq:M3}
\beta_{\text{eff}}(n)= \beta \mathcal{X}_{d}^K\big(\beta_{\text{eff}}(n-1)\big),\quad n=1,2,\dots
\end{equation}
where $n$ indicates the $n$-th stage of the thermomagnetic cascade. In the limit $n\to\infty$, a fixed point solution of Eq.~\eqref{eq:M3} yields the self-consistent equation 
\begin{equation}\label{eq:M4}
m_\mathrm{eq}(\beta) =\mathcal{X}_d\big(\beta m_\mathrm{eq}^K(\beta)\big). 
\end{equation}
In $d=2$, an analytic solution for the thermo-adaptive Ising model can be written thanks to the long-celebrated Onsager-Kauffmann-Yang (OKY) formula for the spontaneous magnetization~\cite{baxter2011onsager, baxter2012onsager}, so that Eq.~\eqref{eq:M4} reads 
\begin{equation}\label{eq:M5}
m_\mathrm{eq}(\beta) =\Big(1-\mathrm{sinh}^{-4}\big(2\beta Jm_\mathrm{eq}^K(\beta) \big)\Big)^{1/8},
\end{equation}
\noindent 
where $J>0$ is the coupling strength. 
For every positive $K$ (not necessarily an integer), the model undergoes mixed-order ferro-paramagnetic (FP) transitions at $T_s(K)\propto K^{-1/8}$ with spinodal scaling $m_\mathrm{eq}(T)-m_\mathrm{eq}(T_s)\propto (T_s-T)^{1/2}$. It is worth to notice that, when simulating the thermo-adaptive Ising model in $d$-dimensional lattices, the global thermal coupling, Eq.~\eqref{eq:M3}, helps smearing out thermal fluctuations, facilitating quenching the system near its spinodal point. \vspace*{+0.25cm}

$\mathrm{M3})$ \textsc{Adaptive contact process}. Consider the $2$-state contact process (CP) illustrated in Fig.\ref{fig:4}\textbf{a}. Nodes change their state from $A\to B$ with rate $\beta\in[0,1]$ and from $B\to A$ with adaptive rate $\Theta^K$, where $K\geq0$ and $[0,1]\ni\Theta=\sum_i a_i/N$ is the fraction of nodes in state $A$ in the network; here, $a_i=1$ if node $i$ is $A$ and $a_i=0$ otherwise. Being a CP, we assume a node changes from $B\to A$ if at least one of its neighbors is $A$. We implement this by measuring the fraction of $A$-neighbors around node $i$---i.e.\ $\Theta_i=\sum_j A_{ij}a_j/k_i$, where $k_i=\sum_j A_{ij}$ and $A=(A_{ij})_{i,j}$ is the network's adjacency matrix---so that $B\to A$ with rate $\Theta_i\Theta^K$. If $K=0$, this CP model undergoes {\em reversible} $A$-to-$B$ phase transitions on $d$-dimensional lattices with dimension $d\geq1$. For $K>0$, instead, the process evolves through a recursive sequence of equilibration steps where, analogously to Eq.~\eqref{eq:M1}, the $B\to A$ rate, denoted here as $\gamma$, evolves adaptively as follows: 
\begin{equation}\label{eq:M6}
\gamma\xmapsto{\Theta(\gamma)}
\gamma_\mathrm{eff}^{(1)}\xmapsto{\Theta(\gamma_\mathrm{eff}^{(1)})} 
\gamma_\mathrm{eff}^{(2)}\xmapsto{\Theta(\gamma_\mathrm{eff}^{(2)})} 
\cdots,\vspace*{-0.1cm}
\end{equation}
for a given rate $\beta\in[0,1]$. By construction, $A$-to-$B$ transitions occur here only for initial condition $a_i(0)=1$ for $i=1,\dots,N$, otherwise the $B$-phase is absorbing. Figures \ref{fig:S4}\textbf{a}--\textbf{c} shows simulations in $3$ networks of same size: (\textbf{a}) complete graph, (\textbf{b}) $d=2$ square lattice and (\textbf{c}) a $d=1$ chain. All models undergo mixed-order $A$-to-$B$ transitions for any $K\geq1$ ($K\geq2$ for the complete graph, see below). Though an exact analytic formula could be derived in the $d=1$ case, we provide below a closed-form solution for the complete graph. The latter can be readily obtained in the fully-mixed approximation~\cite{pastor2015epidemic}, where the discrete variable $a_i\in\{0,1\}$ is replaced by a continuous one $x_i\in[0,1]$ and the adaptive CP, implicitly defined in Eq.~\eqref{eq:M6}, reads now as the set of Langevin equations
\begin{equation}\label{eq:M7}
\dot{x}_i^{(n)}=\beta x_i^{(n)}-\gamma_{\mathrm{eff}}^{(n-1)}\big(1-x_i^{(n)}\big)\sum_j A_{ij}x_j^{(n)},
\end{equation}
where $n=1,2\dots$ denotes the number of feedback iterations. A steady state is reached if $\dot{x}_i^{(n)}=0$ for $i=1,\dots,N$, in which case Eq.~\eqref{eq:M7} yields $x_i=\Theta\sum_j A_{ij}x_j / (\beta + \Theta\sum_j A_{ij}x_j)$; as customary, normalizing the rate governing the $B$-to-$A$ change by $N$, we get the self-consistent solution 
\begin{equation}\label{eq:M8}
\Theta=\frac{\Theta^{K+1}}{\beta + \Theta^{K+1}},\quad K\geq0,
\end{equation}
\noindent 
for the global fraction of $A$-nodes. Notice that $K=0$ yields the classic SIS solution on the complete graph with infection rate $\beta\in[0,1]$ and recovery rate $1/N$. For $K\geq1$, instead, it can be readily check that Eq.~\eqref{eq:M8} undergoes mixed-order $A$-to-$B$ transitions at finite critical rate, $\beta_s$, with the spinodal square-root scaling $\Theta(\beta)-\Theta(\beta_s)\propto (\beta_s-\beta)^{1/2}$. \vspace*{+0.25cm}

$\mathrm{M4})$ \textsc{Random-fuse model}. In a nutshell---for details, see e.g.\ Ref.~\cite{de1985random}---we have simulated the random-fuse model on a $d=2$ lattice (Fig.~\ref{fig:4}\textbf{d}) where each bond is assigned with a current threshold, $I_{c,j}$---sampled uniformly at random from the interval $[-1, 1]$---above which its conductance drops to zero (inset, Fig.~\ref{fig:4}\textbf{d}). We impose a bias current, $I_b$, on the bonds lying at one edge of the lattice and set the voltage to zero (ground) on those lying on the opposite side. We then compute the current flowing through each bond by solving iteratively the Kirchhoff equations with precision $\epsilon=10^{-10}$, as in Ref.~\cite{zapperi1999avalanches}. If a bond fails, we recompute the currents until a steady state is reached or a path of broken bonds spans the lattice, in which case the circuit has reached the fused state, namely where the circuit's conductivity, $G$, drops to zero. The latter occurs abruptly (see Fig.~\ref{fig:S5}\textbf{a}) and with $\phi^3$ spinodal scaling~\cite{zapperi1995self, zapperi1997first, zapperi1997plasticity, zapperi1999avalanches}, that is $G(I_b)-G(I_s)\propto(I_s-I_b)^{1/2}$ and the corresponding quasi-static susceptibility has critical scaling $\chi(I_b)=\partial_{I_b}G(I_b)\propto(I_s-I_b)^{-1/2}$. Indeed, the micro-irreversible nature of the process indicates that the extended avalanches formed prior to the electric breakdown belong to the $\phi^3_2$ universality class. Notice that the long-range character of the avalanches is due to the (long-range) electric interaction; analogous arguments hold for the elastic avalanches formed in the fiber-bundle model~\cite{pradhan2010failure, hansen2015fiber}. In the main text, we focused on the fractal scaling, Eq.~\eqref{eq:3}, by sampling avalanches of fused sites formed during the metastable relaxation (see Fig.~\ref{fig:S5}\textbf{b}) of the circuit from its fully conductive state ($G/N=1$) to the fused one. We did so to averted the difficulty of studying Eq.~\eqref{eq:2} which, in the RF model, features the pathological scaling $I_{s}(L)\propto L/\ln L$ (Fig.~\ref{fig:4}\textbf{c}, inset) of its breakdown threshold~\cite{zapperi1999avalanches}. Details about the thermal-resistively shunted Josephson junction model can be found in Ref.~\cite{bonamassa2023interdependent}. \vspace*{+0.25cm}

$\mathrm{M5})$ \textsc{Relation to other exponents}. 
The two classes of long-range cascades, Eq.~\eqref{eq:1}, complement those predicted by Fisher\! \&\! Berker for first-order (FO) transitions~\cite{fisher1982scaling} and those by Kirkpatrick\! \&\! Thirumalai for random-first-order (rFO) transitions~\cite{gross1985mean, kirkpatrick1987p, kirkpatrick2015colloquium}. Within Kadanoff renormalization scheme~\cite{kadanoff1967static}, one typically starts from the Hankey-Stanley homogeneous form~\cite{hankey1971alternate} of the free-energy density $f(\Delta a, h)\equiv\xi^{-d}f(\xi^{\mathpzc{y}_a}\Delta a,\xi^{\mathpzc{y}_h}h)$, where $h$ is a conjugate field and $\xi$ is the correlation length. Close to the spinodal, the singular part of the order parameter $\psi(a)\equiv\mathcal{O}(a)-\mathcal{O}(a_s)$ scales as $\psi(a)\propto|\Delta a|^{\beta}$ and it is related to $f$ via $\psi(a):=\partial_hf(\Delta a, h)|_{h\equiv0}$; similarly, the susceptibility $\chi:=\partial_{h}^2f(\Delta a, h)|_{h\equiv0}$. Assuming that $f(1,h)|_{h=0}$ is not singular, the above yields the relations $\beta=(d-\mathpzc{y}_h)/\mathpzc{y}_a$ and $\gamma=(2\mathpzc{y}_h-x)/\mathpzc{y}_a$, where $\mathpzc{y}_a=1/\nu$ since $\xi^{\mathpzc{y}_a}\Delta_a=1$ and $\mathpzc{y}_h=\mathrm{D}$. The latter are the classic hyperscaling relations. Now, as shown by Fisher \& Berker, at the coexistence of FO transitions, one finds $\beta=0$ and $\gamma=1$, while in rFO  transitions, $\beta=0$ and $\gamma=2$~\cite{piscitelli2021jamming}. The assumption that hyperscaling holds, yields the set of critical exponents: 

\begin{table}[h]\vspace*{-0.1cm}
	\begin{center}
		\begin{tabular}{c||cc|ccc}
     			 $\,$ & $\,\,\,\beta\,\,\,$ & $\gamma\,\,\,$ & $\,\,\,\nu_d\,\,\,$ & $\mathrm{D}_d\,\,$ & $\eta_d$ \\
      			\hline\hline
      			\textsc{fo}\,\, & $\,\,\,0\,\,\,$ & $1\,\,\,$ & $\,\,\,\nicefrac{1}{d}\,\,\,$ & $d\,\,\,$ & $2-d$\\
			\hline
      			\textsc{rfo}\,\, & $\,\,\,0\,\,\,$ & $2\,\,\,$ & $\,\,\,\nicefrac{2}{d}\,\,\,$ & $d\,\,\,$ & $2-d$\\
  		\end{tabular}\vspace*{-0.5cm}
 	 \end{center}
\end{table}

\noindent 
where, for the anomalous dimension, we assumed that the classic relation $\eta_d=2-\gamma/\nu_d$ holds. We stress that, in the above, no specific notion of symmetry is intended, by contrast with the classes of long-range cascades discussed in the text; moreover, nucleating droplets are expected to be compact in FO and rFO transitions, in contrast with the fractal character, Eq.~\eqref{eq:3}, of long-range $\phi^3_d$ and $\phi^6_d$-cascades. \vspace*{+0.25cm}

$\mathrm{M6})$ \textsc{Cascades in random graphs}. 
The long-range nature of the positive feedback underlying the formation of long-range cascades makes the latter mean-field phenomena from the point of view of the process, regardless the dimension of the underlying structure. In the text, we focused on $d$-dimensional lattices to emphasize this aspect. One might then ask how the scaling relations and exponents in Eqs.~\eqref{eq:1}--\eqref{eq:3} should be interpreted when dealing with long-range cascade processes on random (hyper)graphs. In this case, Eqs.~\eqref{eq:1}--\eqref{eq:3} must be intended with respect to the correlated volume, i.e.\ $|a_s^{\mathrm{av}}(V)-a_s^\infty|\propto V^{-1/\nu^*}$ and $s_a^{\mathrm{max}}(a_s(V))\propto V^{\mathrm{D}^*}$, where 
\begin{equation}\label{eq:M9}
\nu^*=
\begin{cases}
3/2, \\
2; 
\end{cases}
\mathrm{D}^*=
\begin{cases}
2/3&\text{if}\,\,\,\mathds{Z}_2\,\,  \text{broken},\\
3/4&\text{if}\,\,\, \mathds{Z}_2\,\,  \text{conserved}.
\end{cases}
\end{equation}
\noindent 
Here, the correlation volume exponent, $\nu^*$, and the volume fractal dimension, $\mathrm{D}^*$, are related to the correlation length exponent, $\nu$, and fractal dimension, $\mathrm{D}$, via 
\begin{equation}\label{eq:M10}
\nu\equiv \nu^*/\,d_\mathrm{c},\qquad \mathrm{D}\equiv \mathrm{D}^* d_\mathrm{c}, 
\end{equation}
\noindent 
where $d_\mathrm{c}$ is the upper critical dimension of $\phi^3$ or $\phi^6$ spinodals. 
We stress that the exponents of $\phi^3_d$-avalanches in Eq.~\eqref{eq:M10} are consistent with those observed in Ref.~\cite{lee2016hybrid} for interdependent percolation on Erd\H{o}s-R\'enyi graphs, with the important caveat that exponents obtained in Ref.~\cite{lee2016hybrid} do not take into account the stochastic dominance of white noise over finite-size data collapse. 
The above further emphasize that $\phi^3$ and $\phi^6$ long-range cascades are, respectively, manifestations of mean-field percolation~\cite{huang2018critical} and mean-field Ising-like critical phenomena, the latter hinting at a hitherto unknown Coniglio-Klein mapping~\cite{coniglio1980clusters} of $\phi^6$ spinodals onto the critical clusters of some suitable correlated percolation problem.}


\newpage\,
\newpage
\onecolumngrid

\renewcommand\thesection{S\arabic{section}}
\renewcommand\thesubsection{S\arabic{section}.\arabic{subsection}}
\setcounter{section}{0}
\setcounter{equation}{0}
\setcounter{figure}{0}
\setcounter{page}{1}
\renewcommand{\theequation}{S\arabic{equation}}
\renewcommand{\thefigure}{S\arabic{figure}}
\renewcommand{\thepage}{S\arabic{page}}

\vspace*{-1cm}
\section*{\large Supplementary Material}

\begin{figure}[h]
	\includegraphics[width=0.85\linewidth]{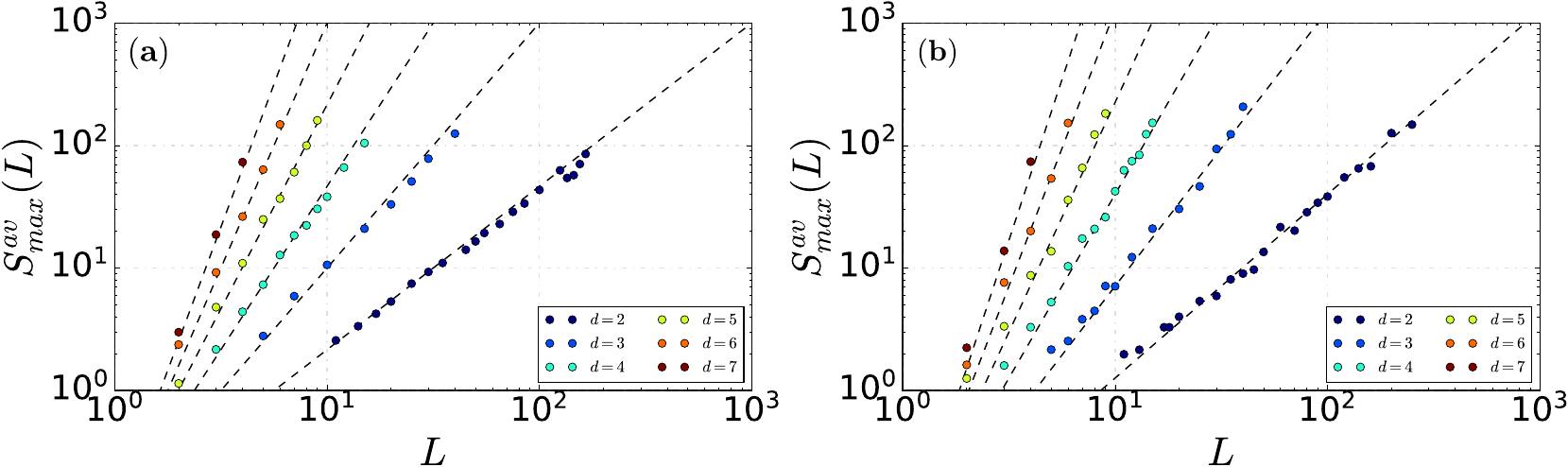}
		\caption{\footnotesize{\textbf{Fractal dimension in thermo-adaptive Ising lattices.} 
		Fractal FSS of the mean mass of large finite avalanches sampled during the metastable (plateau) relaxation from the fully ferromagnetic phase ($\mathcal{M}/N=1$) to the paramagnetic one ($\mathcal{M}/N\sim\mathcal{O}(1/\sqrt{N})$) after quenching the system at $T=T_s(L)+\delta$ with $\delta\leq10^{-3}$, i.e.\ slightly above the mixed-order ferro-paramagnetic transition. Examples of the latter can be found in Refs.~\cite{gross2022fractal, gross2023microscopic}. 
		(\textbf{a}) Results for the $K=1$ thermo-adaptive Ising model on $d$-dimensional lattices with $d=2,\dots,7$ follow the fractal scaling for $\phi^3_d$-cascades, i.e.\ Eq.~\eqref{eq:3} with $\mathrm{D}_d=2d/3$ (dashed lines). 
		(\textbf{b}) Results for the $K=2$ thermo-adaptive Ising model follow instead the fractal scaling for $\phi^6_d$-cascades, i.e.\ Eq.~\eqref{eq:3} with $\mathrm{D}_d=3d/4$ (dashed lines). 
		In \textsc{Kenna} {\em et al}~\cite{berche2012hyperscaling, kenna2014fisher, kenna2017universal, berche2022phase} FSS picture, plots $(\textbf{a}),(\textbf{b})$ show $s^{\mathrm{av}}_{\mathrm{max}}(L)\propto L^{\text{\Koppa}\mathrm{D}}$ where $\text{\Koppa}\equiv d/d_c$, with $(\text{\Koppa},\mathrm{D})=(d/6,4)$ for $\phi^3_d$-cascades and $(\text{\Koppa},\mathrm{D})=(d/4,3)$ for $\phi^6_d$-cascades. 
		}} \label{fig:S1}
\end{figure}

\begin{figure}\vspace*{-5cm}
	\includegraphics[width=\linewidth]{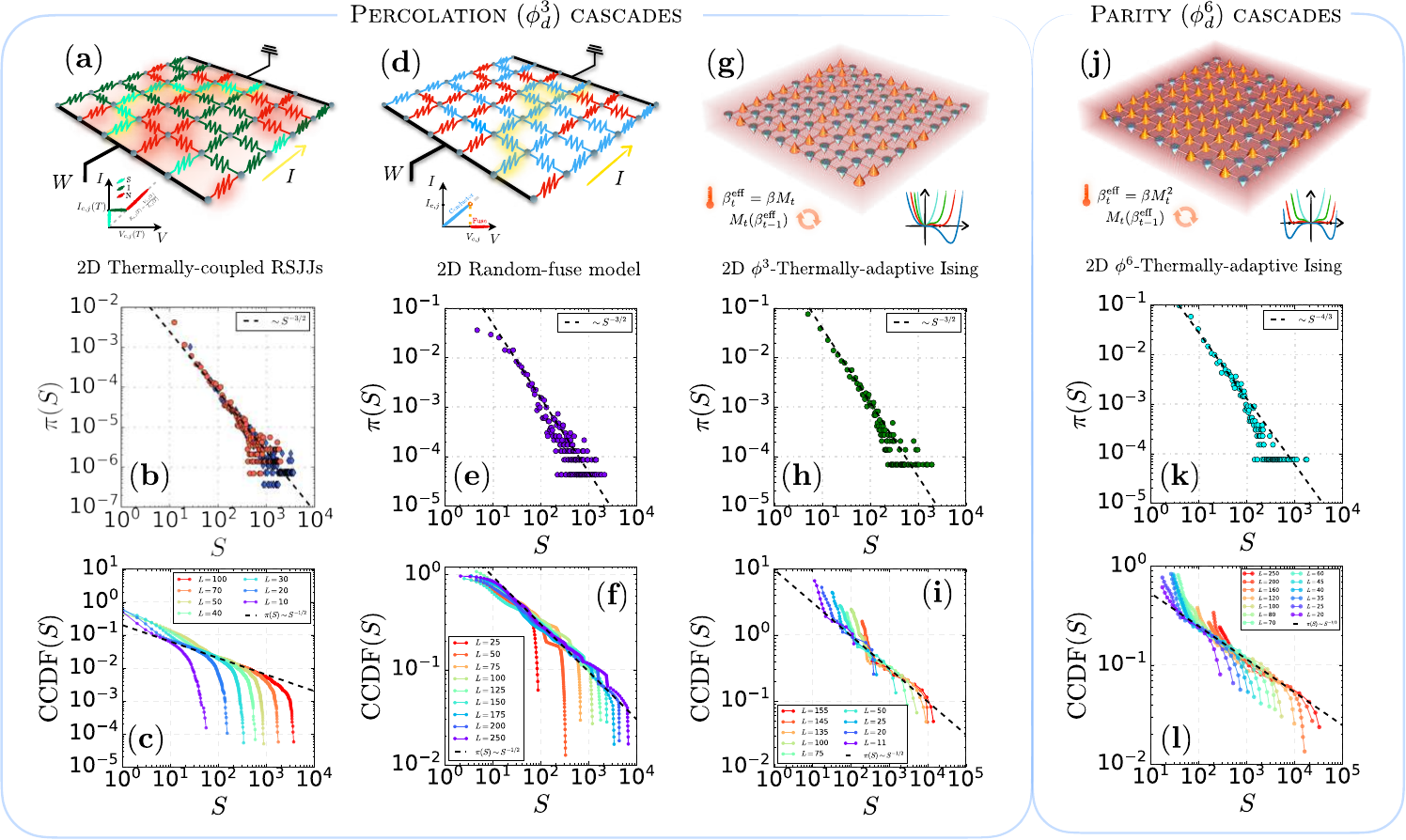}
		\caption{\footnotesize{\textbf{Distribution of finite avalanches sampled during the metastable relaxation above the mixed-order transition. } 
		(\textbf{a}) Thermo-resistively-shunted-Josephson junction model~\cite{bonamassa2023interdependent} in $d=2$. 
		(\textbf{b}) Distribution of the mass of finite avalanches made of bonds switching from the metal to superconducting state (red markers) and viceversa (blue markers); results follow the scaling of $\phi^3_d$-cascades, i.e.\ $\pi(S)\propto S^{-\tau+1}$ with $\tau=5/2$. 
		(\textbf{c}) Complementary cumulative distribution function (CCDF) of $\pi(S)$, i.e.\ $\mathrm{CCDF}(S)\propto S^{-\tau+2}$, for the metal-switching avalanches in (\textbf{b}) highlighting the scaling regime with $\tau=5/2$. 
		(\textbf{d}) Random-fuse (RF) model in $d=2$ (see \textsc{Methods}). 
		(\textbf{e},\textbf{f}) Same of (\textbf{b},\textbf{c}) here for the RF model. 
		(\textbf{g}) Thermo-adaptive Ising model with $K=1$ in $d=2$ (\textsc{Methods}). Notice the asymmetric ($\phi^3$-like) free-energy density (inset). 
		(\textbf{h},\textbf{i}) Same of (\textbf{b},\textbf{c}), here for the $K=1$ thermo-adaptive Ising model in $d=2$. 
		(\textbf{j}) Thermo-adaptive Ising model with $K=2$ in $d=2$ (\textsc{Methods}). Notice the double-well ($\phi^6$-like) free-energy density (inset). 
		(\textbf{k}) Distribution of paramagnetic avalanches following $\phi^6_d$-cascades, i.e.\ $\pi(S)\propto S^{-\tau+1}$ with $\tau=7/3$. 
		(\textbf{l}) CCDF of $\pi(S)$ in (\textbf{k}) highlighting the scaling regime with $\tau=7/3$. 
 		}} \label{fig:S2}
\end{figure}

\begin{figure}
	\includegraphics[width=0.85\linewidth]{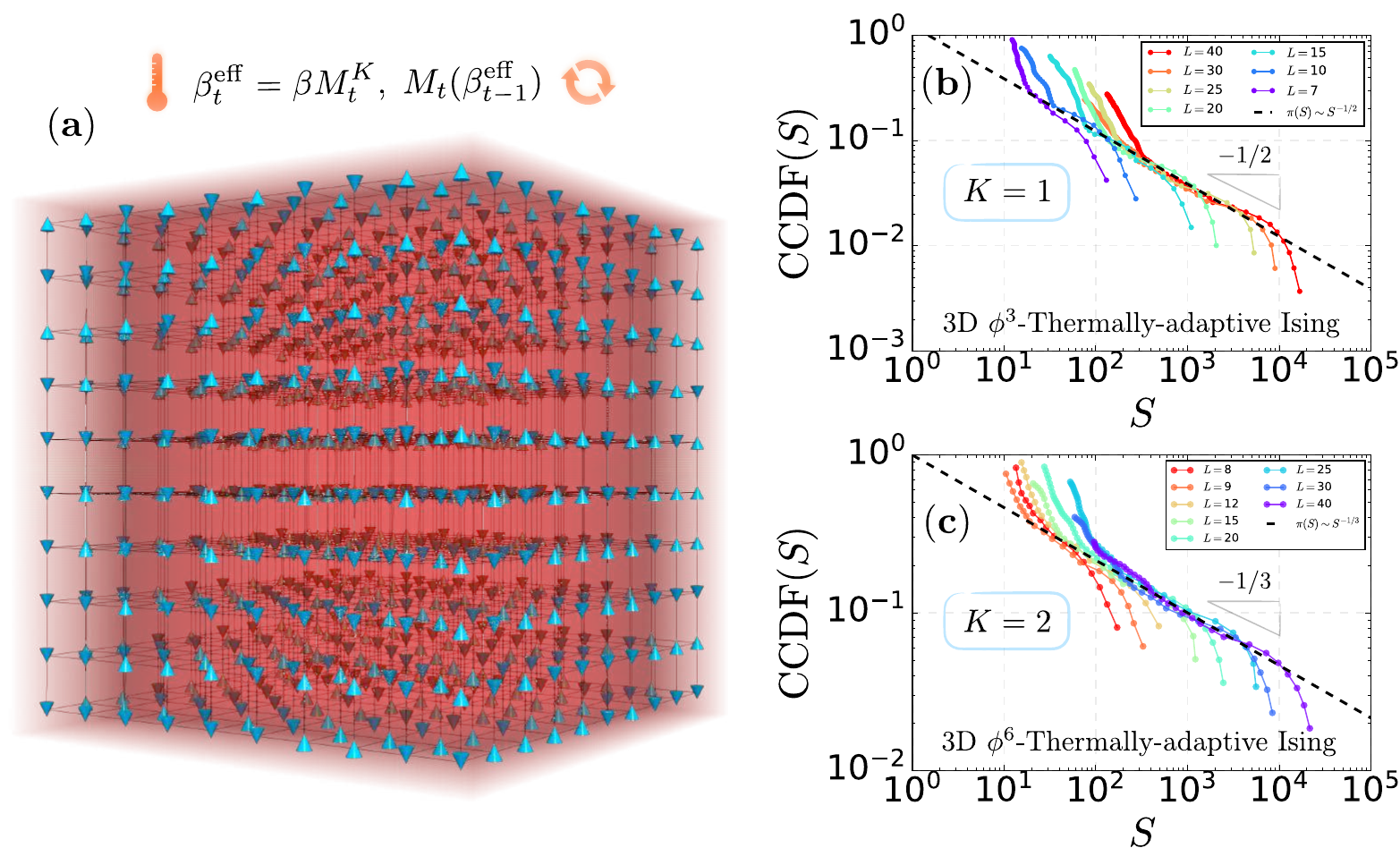}
		\caption{\footnotesize{\textbf{Thermo-adaptive Ising model in $d=3$. } 
		(\textbf{a}) Illustration of the $K$-thermo-adaptive Ising model on a $d=3$ square lattice. Spins, $\sigma_i=\pm1$, interact through ferromagnetic couplings according to the classic Ising Hamiltonian under the influence of a hath-bath with inverse temperature $\beta\equiv 1/T$. The kinetics evolves through an iterative sequence of thermal updates, Eq.~\eqref{eq:M1}, where the feedback is set through the global magnetization reached at the previous iteration (inset). The model undergoes ferro-paramagnetic mixed order transitions for any $d\geq2$ and for every $K\geq1$. 
		(\textbf{b}) Distribution of the mas of finite paramagnetic avalanches sampled during the metastable relaxation of the system from the fully ferromagnetic phase ($\mathcal{M}/N=1$) to the paramagnetic one ($\mathcal{M}/N=\mathcal{O}(1/\sqrt{N})$) at $T=T_s(L)+\delta$ with $\delta\leq10^{-3}$. For $K=1$, the model belongs to the $\phi^3_3$ class and avalanches have power-law distribution, $\tau(S)\propto S^{-\tau+1}$ with $\tau=5/2$; the CCDF associated to $\pi(S)$ highlights this scaling regime. 
		 (\textbf{c}) Same of (\textbf{b}) but with $K=2$, thus for long-range cascades belonging to the $\phi^6_3$ universality class; as expected, $\mathrm{CCDF}(S)\propto S^{-\tau+2}$ with $\tau=7/3$. 
		 Notice the increasing cut-off in both (\textbf{b}) and (\textbf{c})---similarly in Figs.~\ref{fig:S2}\textbf{c},\,\textbf{f},\,\textbf{i},\,\textbf{l}, hinting at the classic expression $\pi(S)\propto S^{-\tau+1}f(S/S_{\mathrm{max}}(L))$ where $f(x)$ is a rapidly decaying function and $S_{\mathrm{max}}(L)\propto L^{1/\sigma\nu_d}$ where $\sigma=1/\nu_d\mathrm{D}_d$, i.e.\ $\sigma=1$ for $\phi^3_d$-cascades and $\sigma=3/2$ for $\phi^6_d$-cascades.}}
		 \label{fig:S3}
\end{figure}

\begin{figure}
	\includegraphics[width=\linewidth]{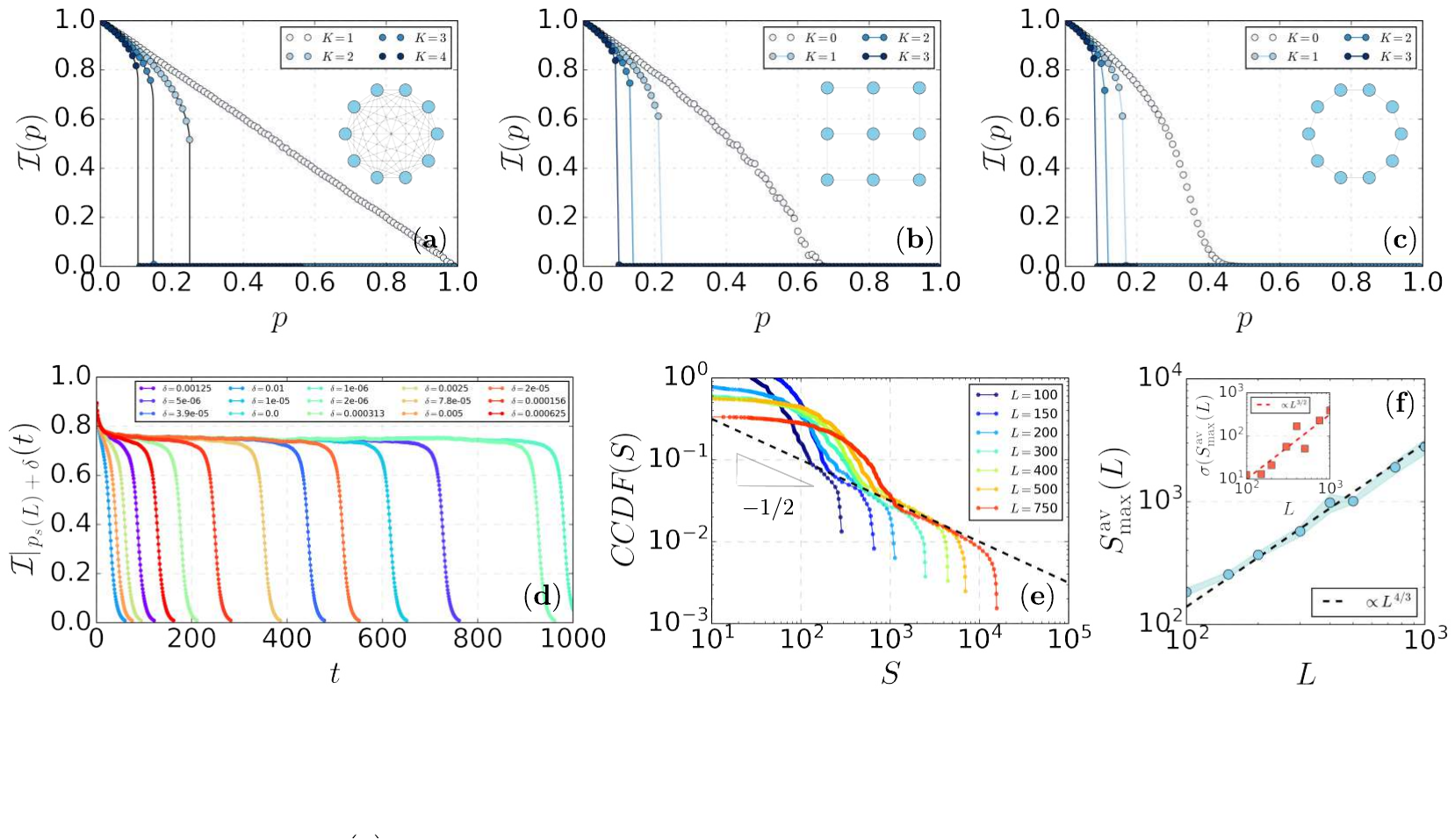}\vspace*{-2cm}
		\caption{\footnotesize{\textbf{Adaptive $2$-state contact process.} As described in $\S$\textsc{M3} in \textsc{Methods}, we have simulated an adaptive contact process (aCP) with $2$ state, where nodes change from $A\to B$ with rate $p\in[0,1]$ and revert from $B\to A$ with adaptive rate $\mathcal{I}^K$, where $\mathcal{I}(p)=\sum_i a_i/N$ is the fraction of nodes with state $A$ in the network. Staring from the initial condition $a_i=1$ for $i=1,2,\dots,N$, the model undergoes mixed-order $A$-to-$B$ transitions for every $K\geq1$ ($K\geq2$ for the fully connected graph, see \textsc{Methods}) and, more intriguingly, for any dimension $d\geq1$ of the underlying lattice. 
		(\textbf{a}) $A$-to-$B$ transitions in fully connected graphs of size $N=10^5$ with $K=1,2,3,4$. Simulations are shown in symbols on top of the curves (full lines) obtained from the self-consistent solution, Eq.~\eqref{eq:M9}. 
		(\textbf{b}) Same of (\textbf{a}) now on a $d=2$ square lattice of size $N=10^6$ with $K=0,1,2,3$. Notice that, differently from the complete graph case, here the $A$-to-$B$ phase transition for $K=0$ occurs at a non-trivial threshold rate $p_c\simeq0.67$. 
		(\textbf{c}) Same of (\textbf{b}) now on a $d=1$ chain of size $N=10^6$ with $K=0,1,2,3$. Notice the continuous, seemingly smeared-out phase transition at $p_c\simeq 0.46$ for $K=0$ and the mixed-order transitions at positive values of $K$.  
		(\textbf{d}) Metastable (plateau) relaxation from the fully $A$-phase ($\mathcal{I}=1$) to the $B$-phase ($\mathcal{I}=0$) obtained by quenching the system at $p=p_s(L)+\delta$ with $\delta\leq10^{-3}$, i.e.\ slightly above the mixed-order $A$-to-$B$ transition. Results are here obtained for the aCP on the $d=2$ square lattice with $L=10^3$ and $K=3$; we stress that the metastable lifetime of the plateau relaxation diverges as we approach the pseudo-spinodal point, $a_s(L)$, i.e.\ one can verify that $\tau(L)\propto (a-a_s(L))^{-1/2}\propto L^{-1/2\nu_d}$ with $\nu_d=3/2d$. 
		(\textbf{e}) CCDF of the $A$-to-$B$ avalanches sampled during the metastable plateau in (\textbf{d}) on$d=2$ lattices with increasing linear lengths; notice the power-law scaling confirming the Fisher's exponent $\tau=5/2$. 
		(\textbf{f}) FSS of the average mass of the largest finite avalanches contributing to the power-law scaling in (\textbf{e}) and following the fractal scaling, Eq.~\eqref{eq:3}, with $D_2=4/3$ (dashed line), expected for $\phi^3_2$-cascades. 
		(Inset) Fluctuations of the largest finite avalanches in (\textbf{f}), showing the phenomenon of stochastic dominance. The latter manifests itself through the ``fractal fluctuation''~\cite{gross2022fractal} scaling $\sigma(S_{max})\propto L{\mathrm{D}_2'}$ where $\mathrm{D}_2'=3d/4$ due to the formal equivalence of critical fluctuations in $\phi^6_d$ cascades and stochastic (gaussian) fluctuations. 
		}} \label{fig:S4}
\end{figure}

\begin{figure}
	\includegraphics[width=0.8\linewidth]{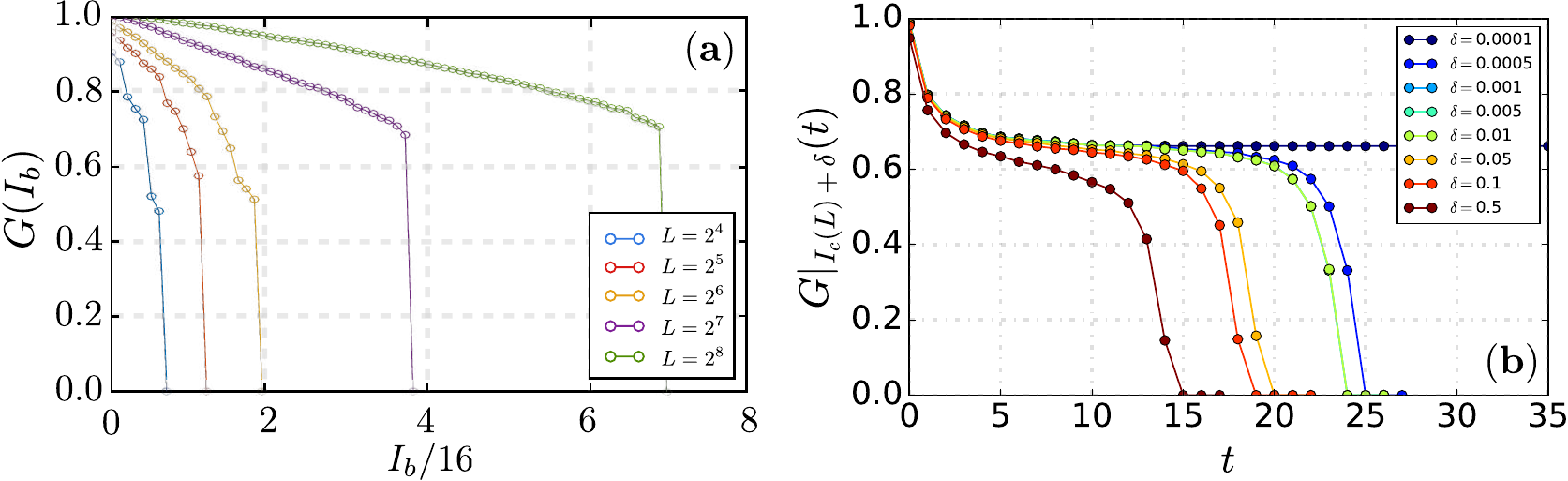}
		\caption{\footnotesize{\textbf{Random-fuse model in $d=2$. } We have solved numerically the Kirchhoff equations characterizing the random-fuse model as briefly described in $\S$\textsc{M5} in the \textsc{Methods}. 
		(\textbf{a}) The model undergoes mixed-order conductor-to-insulator phase transitions at critical values of the bias current which follows the pathological finite-size scaling, $I_s(L)\propto L/\ln L$ (see Fig.~\ref{fig:4}\textbf{e}, inset). As shown in Refs.~\cite{zapperi1997first, zapperi1999avalanches}, the mean conductance, $G$, satisfies the spinodal scaling $G(I_b)-G(I_c)\propto(I_b-I_c)^{1/2}$ with divergent quasi-static susceptibility $\chi=(I_b-I_c)^{-1/2}$. 
		(\textbf{b}) Alike in other mixed-order transitions discussed in the text and in the above, by deep quenching the circuit at $I_b=I_c(L)+\delta$, where here $\delta<0.5$, we observe a metastable (plateau) relaxation from the conductive to the insulator phase, during which a microscopic kinetic process of critical branching underlies the propagation of avalanches of fused bonds throughout the circuit. When a spanning path of broken bonds is formed, the circuit's conductance, $G(t)$, rapidly drops to zero. The largest finite avalanches producing the results in Fig.~\ref{fig:4}\textbf{e} in the main text and in Figs.~\ref{fig:S2}\textbf{e},\,\textbf{f} in the SM have been sampled during this long-lived metastable regime. 
		}} \label{fig:S5}
\end{figure}

\end{document}